%% file: main.tex
\documentclass[sigconf]{acmart}
\makeatletter
\def\@ACM@checkaffil{
    \if@ACM@instpresent\else
    \ClassWarningNoLine{\@classname}{No institution present for an affiliation}%
    \fi
    \if@ACM@citypresent\else
    \ClassWarningNoLine{\@classname}{No city present for an affiliation}%
    \fi
    \if@ACM@countrypresent\else
        \ClassWarningNoLine{\@classname}{No country present for an affiliation}%
    \fi
}
\makeatother

\input{commands}

\renewcommand\footnotetextcopyrightpermission[1]{} 
\AtBeginDocument{%
  }

\begin{document}

\title{Evaluate and Guard the Wisdom of Crowds: Zero Knowledge Proofs for Crowdsourcing Truth Inference}

\author{Xuanming Liu}
\affiliation{%
  \institution{Zhejiang University}
}
\email{hinsliu@zju.edu.cn}

\author{Xinpeng Yang}
\affiliation{%
  \institution{Zhejiang University}
}
\email{yangxinpeng@zju.edu.cn}

\author{Yinghao Wang}
\affiliation{%
  \institution{Zhejiang University}
}
\email{asternight@zju.edu.cn}

\author{Xun Zhang}
\affiliation{%
  \institution{Zhejiang University}
}
\email{22221024@zju.edu.cn}

\author{Xiaohu Yang}
\authornote{The corresponding author.}
\affiliation{%
  \institution{Zhejiang University}
}
\email{yangxh@zju.edu.cn}







\renewcommand{\shortauthors}{authors}

\begin{abstract}
  Crowdsourcing has emerged as a prevalent method for mitigating the risks of correctness and security in outsourced cloud computing. This process involves an aggregator distributing tasks, collecting responses, and aggregating outcomes from multiple data sources. Such an approach harnesses the wisdom of crowds to accomplish complex tasks, enhancing the accuracy of task completion while diminishing the risks associated with the malicious actions of any single entity. However, a critical question arises: How can we ensure that the aggregator performs its role honestly and each contributor’s input is fairly evaluated? In response to this challenge, we introduce a novel protocol termed $\mathsf{zkTI}$. This scheme guarantees both the honest execution of the aggregation process by the aggregator and the fair evaluation of each data source. It innovatively integrates a cryptographic construct known as \textit{zero-knowledge proof} with a category of \textit{truth inference algorithms} for the first time. Under this protocol, the aggregation operates with both correctness and verifiability, while ensuring fair assessment of data source reliability. Experimental results demonstrate the protocol's efficiency and robustness, making it a viable and effective solution in crowdsourcing and cloud computing. 
\end{abstract}

\maketitle

\section{Introduction} \label{sec:intro}
	\def\sectionfolder{sections/}%
	\input{\sectionfolderintroduction.tex}
\section{Preliminaries} \label{sec:pre}
	\def\sectionfolder{sections/}%
	\input{\sectionfolderpreliminaries.tex}%

\section{System overview} \label{sec:overview}
	\def\sectionfolder{sections/}%
	\input{\sectionfolderoverview.tex}%

\section{Zero-knowledge truth inference} \label{sec:main}
	\def\sectionfolder{sections/}%
	\input{\sectionfoldermain-1.tex}%

\section{Transform truth inferences algorithms into circuits} \label{sec:ti}
	\def\sectionfolder{sections/}%
	\input{\sectionfoldermain-2.tex}%

\section{Implementation and evaluations} \label{sec:eval}
	\def\sectionfolder{sections/}%
	\input{\sectionfolderevaluation.tex}%

\section{Conclusion}
	\def\sectionfolder{sections/}%
	\input{\sectionfolderconclusion.tex}%

\bibliographystyle{ACM-Reference-Format}
\bibliography{../../cryptobib/abbrev3, ../../cryptobib/crypto, ./references}

\appendix
	\def\sectionfolder{sections/}%
	\input{\sectionfolderappendix.tex}%

\end{document}

%% file: commands.tex
\usepackage{tikz}
\usetikzlibrary{arrows,shapes}
\usetikzlibrary{shapes.symbols}
\usetikzlibrary{shapes.multipart}
\usetikzlibrary{positioning}
\usetikzlibrary{arrows}
\usetikzlibrary{arrows.spaced}
\usetikzlibrary{matrix} 
\usetikzlibrary{fit}
\usetikzlibrary{calc}
\usepackage{amsthm}
\usepackage{xspace}
\usepackage{framed}
\usepackage{multirow}
\usepackage{graphicx}
\usepackage{subcaption}
\usepackage{enumitem}
\usepackage{bbm}
\usepackage{tcolorbox}

\newtcolorbox[use counter=protocol]{protocolbox}[1][]{
  float,
  label=#1,
  width=\columnwidth,
  boxrule=0.8pt,
  colback=white
}
\usepackage{algorithm}
\usepackage{algorithmic}
\usepackage{setspace}
\usepackage{float}
\usepackage{booktabs}
\usepackage{url}
\usepackage{hyperref}
\usepackage{pifont}
\usepackage{filecontents}

\newtheorem{theorem}{Theorem}
\newtheorem{lemma}{Lemma}
\newtheorem{definition}{Definition}

\newcommand{\brief}[1]{\noindent\textbf{#1}}

\newcommand{\assign}{\ensuremath{:=}\xspace}
\newcommand{\aggregator}{\ensuremath{\mathcal{A}}\xspace}
\newcommand{\dataowner}{\ensuremath{\mathcal{D}}\xspace}
\newcommand{\prover}{\ensuremath{\mathcal{P}}\xspace}
\newcommand{\adversary}{\ensuremath{\mathsf{Adv}}\xspace}
\newcommand{\verifier}{\ensuremath{\mathcal{V}}\xspace}
\newcommand{\zkTI}{\textsf{zkTI}\xspace}
\newcommand{\circuit}{\textsf{C}\xspace}
\newcommand{\worker}{\ensuremath{\mathcal{W}}\xspace}
\newcommand{\task}{\ensuremath{\mathcal{T}}\xspace}
\newcommand{\quality}{\ensuremath{\mathsf{Q}}\xspace}
\newcommand{\prior}{\ensuremath{\eta}\xspace}
\newcommand{\answer}{\ensuremath{\mathsf{V}}\xspace}
\newcommand{\updateTruth}{\ensuremath{\mathsf{update\_truth}}\xspace}
\newcommand{\updateFactor}{\ensuremath{\mathsf{update\_factors}}\xspace}
\newcommand{\CRH}{\ensuremath{\mathsf{CRH}}\xspace}
\newcommand{\ZenCrowd}{\ensuremath{\mathsf{ZC}}\xspace}
\newcommand{\MV}{\ensuremath{\mathsf{MV}}\xspace}

\newcommand{\aux}{\ensuremath{\mathsf{aux}}\xspace}

\newcommand{\FF}        {\mathbb{F}}

%% file: sections/introduction.tex

Many users, finding themselves without the necessary resources to solve problems, must seek help from entities with access to more robust computational capabilities. Outsourcing is a common solution for this issue. However, ensuring that the service provider correctly fulfills its obligations and effectively solves the problem becomes a significant concern. A multitude of literature~\cite{SP:PHGR13, C:GenGenPar10, CCS:FioGenPas14, SP:ZGKPP18, CCS:WJBSTW17} has discussed the security and efficiency issues related to "Verifiable Computing (VC)" and "Verifiable Outsourcing". However, these methods cannot yet be practically applied to very complex tasks.

We propose a different approach: instead of "outsource" the problem to a single entity, we "crowdsource" it to multiple entities. Each entity provides its own answers, and an aggregator then determines the truth of the problem, drawing from this collective wisdom. Crowdsourcing~\cite{tong2020spatial, singer2013pricing} is a popular paradigm for harnessing knowledge from numerous workers. Its advantage lies in the fact that data providers do not need to prove the correctness of their computation to the requester (which previously made up the major cost). As long as they truthfully complete their tasks and contribute their knowledge, their work will be included in an overall consideration (the aggregation process). Furthermore, the collective wisdom of multiple data provider can improve the accuracy of answers and avoid errors caused by single providers.

Despite the potential involvement of malicious data providers, we can generally assume that in real-world scenarios, the majority are honest individuals committed to their responsibilities. They provide accurate solutions and answers, motivated by the prospect of payment. Consequently, it is feasible to extract valid solutions from an array of responses. To address this, we introduce a technique known as \textit{truth inference algorithms}. This category of algorithms is designed to distill the truth of a problem from various candidate answers. It aggregates responses from multiple data providers, deriving a result that epitomizes the "wisdom of crowds." Moreover, these algorithms can consider factors such as each provider's quality and the complexity of the problems, enhancing the accuracy of the outcome. Hence, they are capable of assessing the contributions of each data provider, identifying those who fail to work diligently, and rewarding the diligent ones with more compensations.

However, this scenario introduces another objective: ensuring the correctness of the result while fairly evaluating and safeguarding each data provider's contribution. Our primary concern, therefore, centers on the following question:

\begin{center}
  \textit{How can we ensure that the aggregator correctly and verifiably completes the aggregation process?}
\end{center}

In this context, we define the aggregation as \textit{correctly} executed if the aggregator can accurately infer the truth of the problem upon the wisdom of crowds, and evaluate each data provider's contribution appropriately. Executing \textit{verifiably} means the aggregator must demonstrate to other entities that the process was conducted with integrity, ensuring that results are not incorrect or biased by overestimating certain providers' contributions for personal benefit.

\subsection{Our contributions}

To ensure correctness, we introduce established truth inference algorithms, as referenced in prior works \cite{demartini_zencrowd_2012,li_resolving_2014}. These algorithms are recognized for their effectiveness in extracting the truth from crowdsourced answers across various scenarios. For verifiability, we propose a novel protocol termed \textit{zero-knowledge truth inference} (\zkTI). This protocol leverages a cryptographic primitive known as \textit{zero-knowledge proofs} (ZKPs) to generate a \textit{proof of correct execution} for the truth inference algorithms. To the best of our knowledge, this represents the first application of zero-knowledge proof techniques directly to truth inference algorithms.

Utilizing ZKPs allows us to demonstrate that the aggregator has faithfully executed the aggregation process. This includes receiving results, deducing the truth, and assessing each data provider's contribution, while the verifier can confirm this process without accessing any sensitive data, such as the responses from data providers. Moreover, our protocol's performance significantly surpasses previous efforts \cite{ASIACCS:XLXRZSD20} that employed traditional VC techniques for similar objectives, thanks to our design and the efficient ZKP backends we utilize.

We find that the protocol we propose has a broad range of application scenarios, as realistic use cases given in section \ref{sec:example}. The contributions of our work can be summarized as follows:

\begin{itemize}[nosep, leftmargin=0.3cm, itemindent=0cm]
  \item \textbf{Zero-knowledge proofs for truth inference algorithms.} We present a novel protocol, \textsf{zkTI}, that combines zero-knowledge proof techniques with truth inference algorithms. This integration allows the aggregator to produce a proof of correct execution for the truth inference process. This proof can then be verified by other parties without disclosing any sensitive information. Our protocol ensures that the aggregator uses the responses gathered from workers, and evaluates the quality of each data provider fairly. The need, for aggregator to generate the proof only once,  markedly enhances efficiency.
  \item \textbf{Instantiate algorithms with generic decimal arithmetic.} We have instantiated our proposed protocol with two established truth inference algorithms by converting them into circuits compatible with our ZKP backends. In this process, we designed generic circuits capable of efficiently representing decimal arithmetic in low-level circuits. The performance of our circuits is on par with recent advancements in the field. This advancement not only enhances our protocol with accuracy but also holds promise for application in a variety of other areas, potentially emerging as a subject of independent interest.
  \item \textbf{Implementation and evaluation.} We have conducted a proof-of-concept implementation to validate our protocol. Performance evaluations across different datasets reveal that our approach is at most $4 \times$ more efficient than previous methods~\cite{ASIACCS:XLXRZSD20}, while maintaining algorithmic accuracy. Additionally, our protocol exhibits flexibility, easily adapting to various algorithms and use cases. The source code for our implementation is publicly available (refer to sec. \ref{sec:eval}).
\end{itemize}

\brief{Outline.} Section \ref{sec:pre} provides the preliminaries and essential background knowledge. Section \ref{sec:overview} offers an overview of the system upon which our work is built. The workflow and definition of our protocol are detailed in Section \ref{sec:main}. In Section \ref{sec:ti}, we apply our protocol to two practical algorithms. Our methodology for decimal arithmetic is elaborated in Section \ref{sec:fpc}. Finally, Section \ref{sec:eval} delves into the specifics of our implementation and discusses the outcomes of our experimental evaluations.

\subsection{Related work} \label{sec:related}

\brief{Zero-knowledge proofs and applications in crowdsourcing.} The concept of zero-knowledge proofs (ZKPs) was first introduced by Goldwasser et al. in \cite{STOC:GolMicRac85} and has undergone significant development in recent years. Current research primarily focuses on creating generic ZKP systems for verifying the satisfiability of circuits. The most widely adopted systems are found in references \cite{EC:Groth16, EPRINT:GabWilCio19, EC:BCRSVW19, C:XZZPS19, C:Setty20}. Each of these systems presents different trade-offs in terms of proving time, verification time, and proof size. For our protocol, we have chosen to integrate existing ZKP systems \cite{EC:Groth16, C:Setty20} as our backend.

A number of studies have utilized zero-knowledge proofs (ZKPs) in the realm of crowdsourcing, with a focus on ensuring that worker-provided data meets certain standards. For example, Zhao et al. \cite{zhao2020pace} and Jiang et al. \cite{jiang2021p2ae} have proposed mobile crowdsensing systems (MCS) that use ZKPs to validate the reliability of responses from workers. In a similar vein, Cai et al. \cite{cai2019towards} employed zero-knowledge range proofs in creating a privacy-preserving MCS on a blockchain platform. In this system, blockchain nodes play a crucial role in verifying the authenticity of the data provided by workers. While these studies have a different focus compared to ours, they are complementary to our work, enhancing the broader application of ZKPs in crowdsourcing contexts.

The research by Lu et al. \cite{lu2018zebralancer} presents a crowdsourcing platform that utilizes ZKPs to verify that requesters appropriately compensate workers. While this approach is somewhat aligned with our own, it primarily focuses on the correctness of the workers' rewards and does not address the accuracy of the aggregation process, as it does not employ any truth inference algorithms. Our work, in contrast, advances this concept by integrating truth inference algorithms with ZKPs. This innovative combination ensures not only the precise calculation of rewards for the workers but also the proper execution of the aggregation process by the aggregator.

\brief{Security and privacy of truth inference.} Truth inference algorithms are a key area of research in artificial intelligence and crowdsourcing \cite{zheng_truth_2017, sheng_machine_2019}, primarily focusing on enhancing algorithmic accuracy and broadening their use across diverse scenarios. A notable finding in this field is that the performance of these algorithms varies depending on the scenario \cite{zheng_truth_2017}. 

Our work shifts the emphasis to the security and privacy aspects of these algorithms, an area that has seen growing interest recently \cite{li_efficient_2018, tang_non-interactive_2018, miao2019privacy, ASIACCS:XLXRZSD20, zhang_enabling_2022, wang_privacy-preserving_2022}. Many studies aim to maintain high algorithmic accuracy while protecting the privacy of workers' responses (i.e., their answers) from the aggregator, i.e., an approach known as \textit{privacy-preserving}. These efforts often employ cryptographic techniques like garbled circuits \cite{tang_non-interactive_2018} and differential privacy \cite{li_efficient_2018, wang_privacy-preserving_2022}.  Our work, however, is primarily concerned with the correctness and verifiability of the aggregation process and the fairness in evaluating contributions, making our approach orthogonal to these studies. Nonetheless, the techniques employed in these works could be integrated into our protocol since we do not alter the fundamental process of truth inference, and our protocol is compatible with the algorithm~\cite{li_resolving_2014} commonly used in these studies . 

In our work, the core object is to ensure the verifiability of the truth inference algorithms. Among existing works, \cite{ASIACCS:XLXRZSD20} closely aligns with our objectives. This study utilizes a pairing-based VC technique to render the truth inference algorithm verifiable. However, our experiments suggest that our ZKPs-based method is more efficient than theirs and offers a wider range of application scenarios.

%% file: sections/preliminaries.tex

We have adopted most of the symbols used in truth inference algorithms from \cite{zheng_truth_2017}. The symbol \assign represents equality in a circuit, indicating that the values on both sides of the operator must be equal, while in an algorithm, it denotes an assignment. The Hadamard product is denoted by $\circ$, and $\cdot$ is used to represent a matrix product for matrices and a standard product for integers. "PPT" stands for probabilistic polynomial time. The function $\mathbbm{1}(\cdot, \cdot)$ is an indicator function that outputs 1 when its two inputs are equal and 0 otherwise.

\subsection{Background: crowdsourcing truth inference} \label{sec:bg}

For readers new to the concept of truth inference, we provide some foundational information.

In contexts like crowdsourcing, where information is aggregated from various sources, the presence of low-quality or even malicious contributors is a common challenge. This necessitates the use of a server (aggregator) to collate data from diverse sources and deduce the correct solution to a given problem. This process of aggregating data and deriving the correct answer (the truth) is termed the \textit{truth inference process}, and the algorithm employed for this purpose is the \textit{truth inference algorithm}. This critical issue has garnered considerable attention in the domains of databases (DB) and artificial intelligence (AI), as detailed in the comprehensive review by \cite{zheng_truth_2017}.

\brief{Truth inference.} We formally define the problem herein. For brevity, we consider the problems to be solved as a \textit{task set} of $n$ \textit{tasks}, denoted as: $\task = \{ t_1, t_2, ... , t_n \}$. Participants who contribute their insights (akin to data sources) are referred to as \textit{workers}. Formally, we define $\worker = \{ w_1, w_2, ... , w_m \}$ as a set of $m$ workers. For each task $t_i$, each worker $w_j$ provides an \textit{answer}, denoted as $v_i^j$. Every task $t_i$ in the set $\task$ has a corresponding \textit{ground truth} $v_i^*$, representing the accurate solution. The set of all ground truths is denoted by $\answer^* = \{ v_i^* \}$. Given the set of workers' responses $\answer = \{ v_i^j \}$, the primary objective of a truth inference algorithm is to ascertain the correct answer $v_i^*$ for each task. In certain algorithms, the \textit{quality} \quality of each worker is also evaluated, represented as $q_j$ for worker $j$. These algorithms aim not only to infer correct answers for all tasks but also to accurately assess each worker's quality.

\begin{table}[h]
  \centering
  \begin{tabular}{|c|c|c|c|}
  \hline
  $\worker$: workers  & $\answer$: answers & output                                  & ground truth       \\ \hline
  $w_1$: Alice & $v^1$: 381 m            & \multirow{3}{*}{378 m} & \multirow{3}{*}{381 m} \\ \cline{1-2}
  $w_2$: Bob  & $v^2$: 383 m &  &  \\ \cline{1-2}
  $w_3$: Carl & $v^3$: 370 m &  &  \\ \hline
  \end{tabular}
  \vspace{0.5cm}
  \caption{Example: the height of \textit{the Empire State Building}. The objective is to infer the height of the building using collective responses. Here the output $v^*$ is an average of answers.} \label{tab:1}
\end{table}

\brief{Example.} Consider the task of determining the height of the Empire State Building. Here, the ground truth is the actual height of the building, while the workers represent individuals or entities responding to the query. Their responses constitute the answers. The result of the truth inference is the outcome produced by the algorithm. In Table \ref{tab:1}, a simplistic \textit{average algorithm} might be employed, averaging the responses from the workers. However, this approach does not consider the possibility of unreliable or malicious respondents. Our system implements more sophisticated and accurate algorithms, such as those proposed in \cite{li_resolving_2014} and \cite{demartini_zencrowd_2012}.

\brief{Task types.} As categorized in \cite{zheng_truth_2017}, truth inference tasks are of three types: \textit{decision-making tasks}, \textit{choice tasks}, and \textit{numerical tasks}. Decision-making tasks require binary answers, such as \textit{"Is Argentina the champion of the World Cup?"}. Choice tasks involve selecting an option from a set, like \textit{"Which figure among A, B, C, and D is a cat?"}. These tasks typically require classification of provided objects. If a task presents $l$ possible classifications, they are denoted as $\mathcal{C}={c_1, ..., c_l}$. A choice task with only two options can be considered a decision-making task. Numerical tasks demand numeric responses, such as \textit{"What is the current price of Bitcoin?"}.

\subsection{Zero-knowledge proofs}

To enable an aggregator to demonstrate the correctness of a truth inference algorithm to any verifier who questions the inferred result, we employ the technique of \textit{zero-knowledge proofs}. This work specifically addresses the \textit{circuit-satisfiability} problem. Given a known low-level arithmetic circuit \circuit and \textit{public inputs} $x$, a ZKP system for circuit-satisfiability enables a prover $\prover$ to assure a verifier $\verifier$ of the existence of a \textit{secret witness} $w$ that satisfies the circuit without leaking any information about $w$, denoted as $\circuit(x;w)=1$.

\brief{zkSNARK.} We focus on a class of \textit{zero-knowledge succinct non-interactive argument of knowledge} (zkSNARK) systems. These systems allow the prover to convincingly assert certain statements to the verifier without disclosing any sensitive information. The system comprises a tuple of algorithms $\Pi_\mathsf{ZK}=(\mathsf{Setup, Prove, Verify})$. During the \textsf{Setup} phase, it generates public parameters \textsf{pp}. In the \textsf{Prove} phase, given a circuit \circuit, the prover $\prover$ produces a proof $\pi$ utilizing \textsf{pp} along with public input $x$ and witness $w$. Subsequently, in the \textsf{Verify} phase, the verifier $\verifier$ evaluates the validity of the proof, outputting \textsf{0/1} to indicate whether the proof is accepted or not. The system should exhibit the following properties:

\begin{itemize}[leftmargin=0.3cm, itemindent=0cm]
  \item \textbf{Completeness}. For every \textsf{pp}, valid inputs $x$ and witness $w$, the verifier accepts the proof generated by the prover with the probability \textsf{1}.
  \item \textbf{Knowledge soundness}. For any PPT prover $\prover^*$, there exists a PPT extractor $\mathcal{E}$ that extracts the witness $w^*$ out, that is: 
  
  $\mathcal{E}^{\prover^*}(\mathsf{pp}, x, \pi^*) \rightarrow w^*$. The following relation is $\mathsf{negl{(\lambda)}}$:
  $$
    \Pr[\circuit(x,w^*) \neq 1 \wedge \mathsf{Verify}(x,\pi^*,\mathsf{pp})=1]
  $$
  \item \textbf{Zero-knowledge}. There exists a PPT simulator $\mathcal{S}$ that for any PPT algorithm $\verifier^*$, the output of the simulator is indistinguishable from the real proof. The following relation holds:
  $$
    \mathsf{View}(\verifier^*(\mathsf{pp}, x)) \approx \mathcal{S}^{\verifier^*}(x)
  $$
  \item \textbf{Succinctness}. The proof size $|\pi|$ is of $\mathsf{poly}(\lambda, |x|, \log{|w|})$.
\end{itemize}

\brief{Our zero-knowledge proof backend.}  In our approach, we aim to convert the truth inference algorithm into a low-level arithmetic circuit comprising addition and multiplication gates over a finite field $\mathbb{F}$. With such a circuit in place, we can employ an existing ZKP system as a backend. This backend facilitates the prover in demonstrating that the circuit \circuit is satisfied by certain $x, w$. When considering existing ZKP systems, there are trade-offs concerning prover time, verification time, and proof size. Our goal is to select a ZKP system that can generate proofs as efficiently as possible. Moreover, our proposed application scenarios, which involve the integration of \textit{blockchain} and \textit{smart contracts}, necessitate a non-interactive system with a relatively compact proof size. Given these requirements, we have selected \textsc{Groth16}~\cite{EC:Groth16} and \textsc{Spartan}~\cite{C:Setty20} as the backends for our system's implementation and evaluation. \textsc{Groth16} is the most widely utilized zkSNARK system in the industry, featuring constant-size proofs and optimal verification costs, while \textsc{Spartan} boasts one of the fastest open-source ZKP prover.

It is important to note that both chosen backends operate within the \textit{Rank-1 Constraints System} (R1CS) framework. In this framework, a circuit is represented as a collection of multiplication gates. Consequently, in this study, we utilize the term \textit{gate} to refer to a multiplication gate within a circuit. This serves as a metric for assessing the complexity of circuits.

\subsection{Commitment schemes}

In our approach, we will utilize a \textit{cryptographic commitment scheme} to ensure consistency between the inputs of the truth inference algorithm and the responses provided by the workers. A commitment scheme enables an entity to commit to a specific value while concealing any information about the value until the commitment is opened. This scheme adheres to two fundamental properties: \textit{binding} and \textit{hiding}. The binding property ensures that once committed, the commitment cannot be opened to reveal different values. The hiding property ensures that the commitment does not reveal any information about the value to which it is committed. We denote the commitment process as $\mathsf{Commit}(x; r) \rightarrow \mathsf{com}_x$, where the algorithm takes a value $x$ and, using some randomness $r$, generates the commitment $\mathsf{com}_x$. For simplicity, the randomness may be omitted in certain descriptions.

%% file: sections/overview.tex

In this section, we give an overview of the system model and threat model of our protocol. 

\brief{System model.} Our system encompasses four primary entities:

\begin{itemize}[leftmargin=0.3cm, itemindent=0cm]
  \item \textbf{Data Owner} \dataowner: This entity possesses the problems that require solutions. These problems are distributed to a wide array of workers for resolution, and the truth of these problems is inferred using a truth inference algorithm. The ultimate objective of the \dataowner is to ascertain the truths of all problems.
  \item \textbf{Aggregator (Prover)} \aggregator: Acting as an untrusted entity, the aggregator is responsible for gathering workers' responses to the problems and executing the truth inference algorithm to deduce the truths of these problems. Furthermore, based on the algorithm's outcomes, the aggregator evaluates each worker's performance and allocates corresponding compensation. As \aggregator also operates a ZKP backend to generate proofs, thereby validating the correctness of the algorithm's results and the associated compensations, it is alternatively known as a prover.
  \item \textbf{Workers} \worker: These entities are tasked with executing crowdsourcing tasks and providing solutions to the problems. Their responses are collected by the aggregator \aggregator and utilized as input for the truth inference algorithm. Compensation is awarded based on the quality of their responses.
  \item \textbf{Verifier} \verifier: This entity seeks to verify the accuracy of the truth inference algorithm implemented by the aggregator (prover). In our system, any entity that questions the output can engage as a verifier within the ZKP system and verify the proof provided by \aggregator.
\end{itemize}

\begin{figure}[h]
  \centering
  \includegraphics[width=0.5\textwidth]{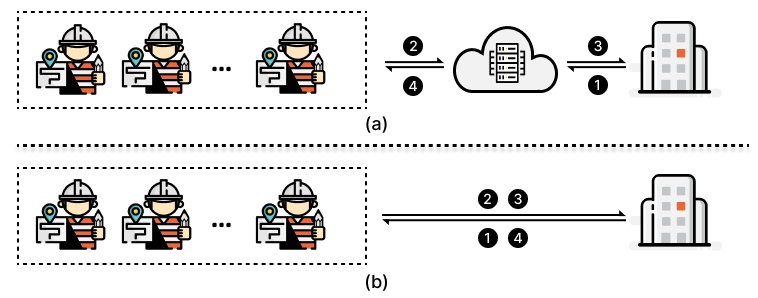}
  \caption{Illustration for two cases of crowdsourcing. In (a) \dataowner crowdsources his problems through a middle entity \aggregator, \aggregator is responsible for the aggregation process. In (b) \dataowner crowdsources the problems to workers directly. It acts both as the data owner and the aggregator. After this process, anyone who doubts about the result can act as a verifier using our \zkTI protocol to check the correctness of the result.}
  \label{fig:fig1}
\end{figure}

It is important to note that in certain scenarios, the data owner and aggregator may be the same entity. For instance, in blockchain oracle services, the service provider might simultaneously function as the data owner (distributing tasks) and the aggregator (collecting responses and inferring truths). We categorize two instances of crowdsourcing and provide a concise description in Figure \ref{fig:fig1}. However, for the purposes of this paper, we focus on the first case, presuming the data owner and aggregator to be distinct entities.

\brief{Threat model.} Our model considers a malicious adversary who may act through \aggregator. After receiving answers from \worker, this adversary might manipulate the algorithm to derive truths and assess \worker' quality in a manner that favors his own interests. Such manipulation could result in \worker not receiving the compensation they merit, thereby harming their interests. Additionally, \dataowner may be provided with incorrect results. We further assume that the adversary could corrupt a portion of \worker, leading them to submit inaccurate answers, thereby disrupting the process of inferring correct responses. Finally, we assume a curious \verifier who may attempt to extract information from the proof generated by \aggregator.

\brief{System goals.} The objectives of our protocol are threefold: (i) \verifier should only accept a proof if \aggregator has correctly executed the designated truth inference algorithm and presented valid outputs, which include both the truths to the problems and an assessment of the workers' quality. (ii) Any attempts by corrupted workers to provide incorrect answers with the intent of disrupting the aggregation process should be detectable. (iii) \verifier should not gain any information from the verification process, except for what is derived from their own input. In our scope, \worker should not learn the truth of tasks, while \dataowner should not learn the answers provided by \worker.

%% file: sections/main-1.tex

At a high level, crowdsourcing allows data owners (requesters) to obtain higher quality data, avoiding trust issues in the outsourcing computation process. The truth inference algorithm plays a crucial role in this process, aiding in the aggregation of collected responses and synthesizing high-quality answers. A zero-knowledge truth inference (\zkTI) is a service that
enables the crowdsourcing process to run in a secure and verifiable fashion.
Below, we outline the design of the crowdsourcing workflow and elucidate how \zkTI facilitates this process, detailing the protocol more comprehensively.

\brief{Crowdsourcing workflow.} Fig. \ref{fig:fig1} depicts the crowdsourcing workflow, facilitated by the \zkTI protocol. We illustrate this using subgraph (a) as an example, though subgraph (b) is also supported. Consider a data owner \dataowner with a set of problems $\task$ requiring solutions. In the crowdsourcing workflow, \dataowner initially delegates these tasks to an aggregator \aggregator (Step \ding{182} in Fig. \ref{fig:fig1}). \aggregator then distributes these tasks among numerous workers $\worker$ (Step \ding{183}). Each worker $w_i \in \worker$ is tasked with completing a subset of tasks $\task_i \subseteq \task$ and providing answers. Upon receiving the answers, \aggregator employs a truth inference algorithm $f$ to deduce the truth of each task. Subsequently, \aggregator presents the inferred truths to the data owner (Step \ding{184}), who utilizes them to address their problems. Moreover, the algorithm $f$ estimates the quality $\quality$ of each worker, which \aggregator uses as a basis to evaluate performance and allocate compensation accordingly (Step \ding{185}). Higher-rated workers receive greater compensation, and vice versa. By the combination of a truth inference algorithm, a low-quality or even malicious worker cannot obtain a higher rating and receive more compensation than they merit. 

Following this workflow, any entity questioning the algorithm's integrity can engage in a \zkTI protocol with \aggregator. Here, \aggregator is required to provide proofs affirming the proper use of collected data as inputs and the correct execution of the algorithm to produce the results. For instance, a worker disputing their compensation or a data owner concerned about problem resolution may assume the verifier's role to examine proofs offered by the aggregator (prover).

To persuade \verifier of the honest execution of the algorithm, we begin by transforming the truth inference algorithm $f$ into a low-level arithmetic circuit \circuit comprising addition and multiplication operations. The details of this transformation will be discussed in section \ref{sec:ti}. With this circuit in place, \aggregator can utilize an existing ZKP backend to generate a proof $\pi$, demonstrating that the circuit satisfies both the collected data (inputs) and the algorithm's outputs.

\subsection{Prove once for multiple verifiers} \label{sec:po}

Once the algorithm $f$ is determined and transformed into a circuit \circuit, the aggregator \aggregator (acting as a prover) must demonstrate that \circuit is satisfied given the inputs and outputs. Formally, \aggregator needs to prove $\circuit(\answer, \answer^*, \quality)=1$, where $\answer$ represents the responses collected from workers, $\answer^*$ is the inferred truth of each task, and $\quality$ is the inferred quality of each worker. It becomes essential to distinguish which components are the (secret) witness $w$ and which are the public inputs $x$.

\brief{Problem: multiple verifiers with different inputs.} A key challenge arises when multiple entities (workers and data owner), each with their unique secret inputs $w$, seek to verify the aggregator's execution of the algorithm. These verifiers should not gain any information beyond their respective $w$. For example, if a verifier is a worker-$j$, possessing only a subset of responses $\answer_j = \{v_i^j\}_{i \in [n]}$, they might require \aggregator to validate the correctness of their answer quality assessment $q_j$, which determines their compensation. The system goal compels \aggregator to establish this relationship without revealing information about other workers' answers $\answer - \answer_j$ or the inferred truths $\answer^*$. Conversely, a data owner \dataowner may demand proof that $\answer^*$ is accurate, without gaining access to \answer. This is reasonable as in some scenarios, the data owner may not be authorized to access the collected responses. Naively, the above demands require \aggregator to carefully select the witness and public inputs and generate proofs multiple times for different verifiers, necessitating continuous online presence and repeated engagement in the proof generation process. This imposes a considerable burden on \aggregator and underscores the need for a system design that minimizes the aggregator's workload while still satisfying the diverse verification requirements of multiple entities.
 
\brief{Prove in one time.} Existing works in the literature~\cite{USENIX:ZXHSZ22, AC:YanWan22} address ZKP systems involving multiple verifiers. However, these works either do not suit our specific context or require the incorporation of other complex cryptographic primitives, indicating the necessity for a customized solution that conforms to our system's unique requirements without introducing excessive complexity or deviating from our established framework.

Our approach is influenced by the insights from~\cite{CCS:CamFioQue19}. The primary objective is twofold: firstly, to enable the prover to prove once, while allowing multiple verifiers, each with different inputs, to conduct their respective verifications. This process must ensure that verifiers do not acquire any information other than their own inputs during the proof. Secondly, it is essential to maintain data consistency, assuring the verifier that the inputs to the algorithm and the answers collected from the workers are coherent. These inspire us to introduce a cryptographic commitment scheme. Each worker is required to commit to their own responses $\answer_j$ using $\mathsf{com}_{\answer_j} = \mathsf{Commit}(\answer_j)$ and subsequently reveal these commitments. Similarly, the data owner \dataowner can commit to the predicted truths $\answer^*$, resulting in $\mathsf{com}_{\answer^*}$. Following this, \aggregator is tasked with opening all commitments within circuit \circuit, a feasible task given \aggregator's knowledge of $\answer$ and $\answer^*$. If the commitment scheme is binding, \aggregator can assert to the verifier that "I have genuinely used the collected answers $\answer$ as input and derived the outputs $\answer^*$ and \quality, where $\mathsf{com}_{\answer_j}$ corresponds to $\answer_j$ and $\mathsf{com_{\answer^*}}$ to $\answer^*$". This strategy ensures that each worker-$j$ does not access information about other workers' responses from the proof, while maintaining data consistency. It also facilitates the participation of \dataowner in the verification process, ensuring output integrity and revealing only the commitments. 

In conclusion, the circuit \circuit, which \aggregator is required to prove, encompasses both the truth inference algorithm $f$ and the process of opening all commitments. The relationship within this framework can be formulated as follows:
\begin{equation}
  \circuit(\{\mathsf{com}_{\answer_j}\}_{j\in[m]}, \mathsf{com}_{\answer^*}, \quality; \answer, \answer^*) = 1
\end{equation}
Consequently, this allows \aggregator to provide a comprehensive proof, for only once, to validate the correctness of the entire process to any verifier.
It is important to note that while the quality of workers (\quality) is considered a public input here, it can also be encapsulated as a witness using commitments. 

In our system, the commitment scheme is implemented using a collision-resistant hash function, as delineated in~\cite{USENIX:GKRRS21}. Detailed information regarding this implementation is provided in section \ref{sec:eval}.

\subsection{Zero-knowledge truth inference protocol} \label{sec:zkTI}

Building on the concepts discussed earlier, we delineate the definition of the zero-knowledge truth inference (\zkTI) protocol. Formally, let $\mathbb{F}$ represent a finite field, $\task$ denote $n$ tasks designated for crowdsourcing, $\answer$ signify the answers provided by $m$ workers $\worker$, and $\quality$ indicate the quality attributed to each worker. The \zkTI protocol is characterized by the following algorithms:

\begin{itemize}[leftmargin=0.3cm, itemindent=0cm]
  \item $\mathsf{pp} \leftarrow \mathsf{zkTI.Setup(1^\lambda)}$: Given the security parameter, generate the parameters needed.
  \item $\{\mathsf{com}_{\answer_j}\}_{j\in[m]}, \mathsf{com}_{\answer^*} \leftarrow \mathsf{zkTI.Commit}(\{\answer_j\}_{j\in[m]}, \answer^*)$: \worker and \dataowner commit their own inputs.
  \item $\pi \leftarrow \mathsf{zkTI.Prove}(\{\mathsf{com}_{\answer_j}\}_{j\in[m]}, \mathsf{com}_{\answer^*}, \answer, \answer^*, \quality, f, \mathsf{pp})$: \aggregator generates a proof $\pi$ to prove that $\answer^*, \quality$ is the output of $f$ on the inputs and the opening of all commitments is valid.
  \item $\{\mathsf{0,1}\} \leftarrow \mathsf{zkTI.Verify}(\{\mathsf{com}_{\answer_j}\}_{j\in[m]}, \mathsf{com}_{\answer^*}, \quality, \pi, \mathsf{pp})$: Any verifier can validate the proof $\pi$ with the public parameters $\mathsf{pp}$ and the public inputs. He outputs 1 if the proof is valid, otherwise outputs 0.
\end{itemize}

The protocol should have the properties of completeness, soundness and zero-knowledge as the generic zero-knowledge proofs. We give the formal definition in Appendix \ref{app:D}.

\subsection{Example applications} \label{sec:example}

In section \ref{sec:ti}, we will delve into the selection of the truth inference algorithm $f$ and the detailed design of the circuit \circuit. Before that, we present some exemplary use cases of the proposed protocol.

\brief{Data Annotation in AI/ML.} Data is a cornerstone in machine learning (ML) and artificial intelligence (AI), with crowdsourcing for data annotation being a common practice to obtain labeled datasets for model training.

Consider Scale.AI~\cite{noauthor_accelerate_nodate}, a prominent data annotation company. It employs two primary annotation methods. The first method~\cite{noauthor_rapid_nodate} involves the company \aggregator handling datasets from clients \dataowner for annotation, distributing them among various annotators \worker, and subsequently aggregating and returning the annotated data to \dataowner. The second method~\cite{noauthor_studio_nodate} enables organizations to form their own annotation teams, assign tasks, and compensate them accordingly.

In either method, the objectives are accurate data annotation and equitable evaluation and remuneration of annotators. Our \zkTI protocol seamlessly fits into these scenarios. In the first method, \aggregator can utilize the \zkTI protocol to compile annotations from \worker and provide \dataowner with results and proof for verification. In the second method, corresponding to case 2 in Fig. \ref{fig:fig1}, \dataowner can directly aggregate responses and quantify each worker's contribution. Workers can then validate the fairness of their evaluation and compensation using our protocol.

\brief{Blockchain Oracles.} Blockchain oracles~\cite{caldarelli2021blockchain} act as bridges, fetching off-chain data for on-chain smart contracts, ranging from stock prices to weather forecasts or sports results.

Typically, a blockchain oracle involves a server gathering and publishing data to a smart contract. However, centralized approaches have vulnerabilities, such as single-point failures or the risk of incorrect data dissemination~\cite{noauthor_oracle_nodate}.

Our protocol is advantageous for developing decentralized oracles. It facilitates the integration of multiple data sources (\worker), including authoritative bodies or individuals. When a smart contract (\dataowner) needs data, a server \aggregator collects potential responses from these sources, processes them via our workflow, and supplies the results with proof to the smart contract, which then acts as a verifier. Successful verification indicates honest data aggregation by the server. Furthermore, the server can compensate data contributors based on their assessed input, as determined by the algorithm, which is also verifiable by the smart contract.

%% file: sections/main-2.tex

At this moment, having outlined the \zkTI protocol and its application in the crowdsourcing process, we will now focus on the intricacies of transforming the truth inference algorithm into an arithmetic circuit and executing a ZKP backend on it. This section aims to select suitable truth inference algorithms and delineate the development of a versatile circuit capable of verifying each computational step. We will conclude by presenting a comprehensive construction of the \zkTI protocol within our system, ensuring it aligns with our system’s objectives.

\subsection{Algorithm Representation} \label{sec:algos}

The challenge lies in converting the selected algorithm $f$ into a circuit. This conversion entails deconstructing the algorithm into fundamental operations representable in an arithmetic circuit, like additions, multiplications, and logical functions. Each step in the algorithm corresponds to one or more components in the circuit. The algorithm should be conducive to circuit representation for ease of integration.

Efficiency is paramount in our selection of the algorithm. The complexity of the circuit \circuit should ideally be a function of the number of tasks and the number of workers involved in the crowdsourcing, rather than the complexity or size of the tasks themselves. This criterion is vital for ensuring the scalability and adaptability of our protocol to various crowdsourcing scenarios. We will now introduce our chosen algorithm and explain its suitability for our purposes.

\brief{A classical framework of truth inference.} In our system, it is essential to select an algorithm that acknowledges the quality of workers and accurately deduces the truth of the problems. Recent research \cite{wu2021data} has underscored the significant influence of worker quality on the accuracy of truth inference algorithms. This insight implies the necessity of placing greater trust in responses from high-quality workers to more precisely infer the actual answers to questions. Reflecting this understanding, most contemporary approaches typically adopt a two-stage framework (illustrated in Alg. \ref{alg:tif}). Initially, the quality of workers is established using prior knowledge, which is then iteratively refined throughout the algorithm's execution.

\begin{algorithm}
  \setstretch{1.05}
	\caption{Truth inference framework $f$}
  \renewcommand{\algorithmicrequire}{\textbf{Input:}}
  \renewcommand{\algorithmicensure}{\textbf{Output:}}
	\begin{algorithmic}[1]\label{alg:tif}
    \REQUIRE {workers' answers \answer and prior factors \prior}
    \ENSURE {inferred the truth $\answer^*$ and workers' quality \quality}
    \STATE $\mathsf{iter \assign 0}$
    \WHILE {true}
      \STATE $\mathsf{iter} \assign \mathsf{iter} + 1$

      \STATE // Update the inferred truth $\answer^*$
      \STATE $\answer^* \assign \updateTruth(\answer, \prior)$

      \STATE // Update worker's quality
      \STATE $\quality \assign \updateFactor(\answer, \answer^*)$

      // Break if the algorithm converges or reaches the maximum iteration
    \ENDWHILE
    \RETURN $\answer^*, \quality$
	\end{algorithmic}  
\end{algorithm}

The framework works as follows:  initially, we establish the quality of workers and other variables (such as task difficulties) based on \textit{prior knowledge}. For instance, a worker's past performance in crowdsourcing can be an indicator of their current answer quality. In our approach, these prior factors are denoted as \prior and considered as public inputs. Subsequently, the algorithm iteratively updates the quality of workers and the truth of tasks until convergence is achieved or a maximum iteration limit is reached. Convergence is defined as the state where the predicted truths of the problems and the quality assessments of the workers stabilize, and both \aggregator and \dataowner acknowledge and accept these results. We modify the relation to be proven as $\answer^*, \quality = f(\answer, \prior)$. In this work, we focus on that during each algorithm iteration, \aggregator must demonstrate to other participants that the algorithm has been executed truthfully. The protocol can be executed multiple times to iteratively process the algorithm, ensuring the trustworthiness of the overall procedure.

\brief{Instantiation.} \label{sec:instance} Drawing from this two-stage framework, various truth inference algorithms have been developed and successfully applied across different fields. For our system, we require an algorithm that is not only highly accurate and adaptable to diverse scenarios but also conveniently representable within a circuit \circuit. We opt for two existing algorithms, \cite{li_resolving_2014} and \cite{demartini_zencrowd_2012}, for our protocol instantiation.

The first algorithm, \CRH~\cite{li_resolving_2014}, is frequently utilized in literature concerning the security and privacy of truth inference~\cite{ASIACCS:XLXRZSD20, li_efficient_2018, zheng_truth_2017} due to its ease of circuit representation and proven accuracy in practical implementations. With minor adjustments, \CRH can be represented as follows:

\begin{itemize}[nosep, leftmargin=0.3cm, itemindent=0cm]
  \item \textbf{Update truth:} Given the quality $q_j$ of each worker-$j$, the inferred truth $v_i^*$ of each task-$i$ is updated in a weight-average manner:
  \begin{equation}\label{eq:crh-1}
    v_i^* = \frac{\sum_{j=1}^{m} q_j * v_{i}^j}{\sum_{j=1}^{m} q_j}
  \end{equation}  
  \item \textbf{Update quality:} Given $v_i^*$ as the inferred truth of task-$i$, each worker's quality $q_j$ is updated as:
  \begin{equation}\label{eq:crh-2}
    q_j = \log(\frac{\sum_{j'=1}^m\sum_{i=1}^n{d(v_i^{j'},v_i^*)}}{\sum_{i=1}^n{d(v_i^j,v_i^*)}})
  \end{equation}
\end{itemize}
In the context of our protocol, $d$ represents a distance function used to calculate the discrepancy between a provided answer and the inferred truth. For decision-making tasks, this function can be implemented using the indicator function $\mathbbm{1} (\cdot, \cdot)$. The indicator function can be represented in circuit using selector operations. Additionally, the logarithm operation, being a public function, can be deferred to the verifier as he can calculate it on his own. Hence, it is apparent that each step of the algorithm can be seamlessly transformed into operations within a circuit, enabling the entire algorithm to be reformulated in circuit form.

Furthermore, we incorporate the \ZenCrowd algorithm~\cite{demartini_zencrowd_2012} into our protocol. \ZenCrowd utilizes a probability-based approach for choice-making tasks. In the \updateTruth stage of the algorithm, it calculates the probability of each choice $c_k$ being the correct answer using a specific formula:
\begin{equation}\label{eq:zencrowd}
  \Pr(v_i^* = c_k) = \prod_{j=1}^{m}(q_j)^{\mathbbm{1} (v_i^j, c_k)} \cdot (1-q_j)^{1 - \mathbbm{1} (v_i^j, c_k)}
\end{equation}
This step is followed by a normalization process, and the inferred truth for a task is identified as the choice with the highest probability. These steps can be conveniently represented in a circuit as well. 

The problem of the above instantiation is that we need to introduce decimal arithmetic. Considering that the algorithms necessitate division and that our inputs, along with certain parameters, will be represented as decimals, the circuit must facilitate arithmetic operations with decimal numbers. As shown in Equation \ref{eq:zencrowd}, this calculation involves $m$ successive decimal multiplications, with the total number of multiplications required being $O(n \cdot m \cdot l)$, where $n, m, l$ denote the number of tasks, workers, and choices, respectively. We encounter a challenge with these consecutive multiplications, as they do not progress smoothly using standard fixed-point multiplication for decimal numbers. We address this issue in section \ref{sec:fpc} by introducing generic designed decimal arithmetic circuits that facilitates integration with a ZKP backend.
 
\subsection{Generic circuits for decimal arithmetic} \label{sec:fpc}

From our selection of algorithms, it becomes clear that the core computations consist of decimal multiplication, addition, and division operations. This section is dedicated to designing generic circuits that accommodate these operations, particularly focusing on arithmetic involving decimals.

Existing approaches either interpret decimal numbers as fixed-points within a significantly large finite field~\cite{ASIACCS:XLXRZSD20}, or perform full-fledged floating-point computations within a circuit~\cite{USENIX:WYXKW21}. The fixed-point method faces limitations in handling a series of consecutive multiplications, as the magnitude of the resultant numbers exponentially increases with each multiplication, potentially surpassing the range of the finite field. This limitation becomes apparent in our system, as indicated by Equation \ref{eq:zencrowd}, which involves numerous successive multiplication operations. On the other hand, complete floating-point computation incurs considerable overhead. For example, \cite{USENIX:WYXKW21} reports that about 8,000 gates are required to validate a multiplication of two IEEE-754 floating-point numbers~\cite{noauthor_ieee_2023}, rendering this approach impractical for widespread use. Moreover, accurate decimal arithmetic is also vital in other domains such as finance, mathematics, and machine learning.

In light of these considerations, we propose the development of generic circuits that support decimal arithmetic within a finite field. This approach aims to strike a balance between computational feasibility and the practical requirements of diverse applications where decimal calculations are indispensable.

\brief{Floating-points.} Our start point is floating-points. A floating-point number can be represented as $v = s \cdot 2^{e-e_0}$, where $e$ is the \textit{exponent}, $e_0$ is a fixed shift, and the $|s| \in [2^{w-1}, 2^w)$ is a $w$\textit{-bits} integer where $w$ decides the precision. One thing worth noting is that the \textit{most-significant} bit of $s$ is always $1$ due to the $normalization$ operation. For example, for a 32-bit floating-point number, $w=23$ and $e$ is an 8-bits integer. For simplicity, in this paper we assume $s$ is a positive number, and we omit the fixed $e_0$ in the exponent during the description. So a decimal number $v$ can be represented as a floating-point using a pair $(s, e)$. 

\brief{Our design.} In contrast to the approach proposed in \cite{USENIX:WYXKW21} for proving the correctness of floating-point computations in a bit-by-bit manner, which incurs significant overhead, we adopt a more holistic perspective. Specifically, when verifying whether $c = (s_c, e_c)$ (an input provided by the prover) is indeed a valid result of a computation $g(a,b)$, we direct the prover to demonstrate that $c$ has been accurately computed according to a specified computation process.

For example, to understand this in the context of decimal multiplication, consider computing the product of decimals $a$ and $b$ in plain mathematics. Initially, we compute $\hat{s_c} = s_a \cdot s_b$ and then normalize this result to $\tilde{s_c} = \hat{s_c} \cdot 2^{-\theta}$, ensuring that the most-significant bit of $\tilde{s_c}$ is 1, where $\theta$ is the normalization factor (equivalent to right-shifting $\hat{s_c}$ by $\theta$ times). Consequently, the exponent is calculated as $e_c = e_a + e_b + \theta$. However, it's important to note that $\tilde{s_c}$ may not be an integer, which necessitates a process termed \textit{rounding}. It involves reducing $\tilde{s_c}$ to its floor value, resulting in $s_c = \lfloor \tilde{s_c} \rfloor$. 

Our objective is to encapsulate the entire computation process, including the rounding operation, within a circuit. While most operations above are deterministic and can be straightforwardly represented in a circuit, the rounding operation poses a challenge due to its non-deterministic nature. To validate the correctness of rounding, we adopt the \textit{relative error} model as proposed in \cite{CCS:GJJZ22}. This model allows us to demonstrate the accuracy of the rounding operation by validating the following relationship:
\begin{equation}\label{eq:rel_mul}
  \tilde{s_c} - s_c \leq \delta \cdot \tilde{s_c}
\end{equation}  
Here, $\delta$ represents a commonly agreed input known as the \textit{relative error} or precision. The verifier will deem the given $s_c$ as a correct computation result if the aforementioned relation is satisfied. Specifically, for $w$-bit floating-point numbers, the precision $\delta$ can be set to $2^{-(w-1)}$. Adhering to this framework, the multiplication of two floating-point numbers can be effectively constrained by the following relationships:
\begin{equation}\label{eq:our_mul}
  \begin{split}  
    s_a \cdot s_b \cdot 2^{-\theta} - s_c &\leq \delta(s_a \cdot s_b  \cdot 2^{-\theta}) \iff \\ 
    s_a \cdot s_b - s_c \cdot 2^\theta &\leq \delta (s_a \cdot s_b), \\
    e_c - \theta &= e_a + e_b
  \end{split}
\end{equation}  

The same idea applies for other operations. When performing division of $a$ by $b$, we first compute the division $\hat{s_c} = \frac{s_a}{s_b}$. After this calculation, normalization of $\hat{s_c}$ is applied to yield $\tilde{s_c} = \hat{s_c} \cdot 2^{\theta}$, which is equivalent to left-shifting $\hat{s_c}$ by $\theta$ times. The exponent part of the result conforms to the equation $e_c + \theta = e_a - e_b$. Finally, similar to the multiplication operation, we round $\tilde{s_c}$ down to the nearest floor value, thereby obtaining $s_c$. Thus, for division, the constraints can be formulated as follows:
\begin{equation}\label{eq:our_div}
  \begin{split}  
    s_a \cdot 2^\theta - s_c \cdot s_b &\leq \delta (s_a \cdot 2^\theta),  \\
    e_c + \theta &= e_a - e_b
  \end{split}
\end{equation}  

For addition operations, handling differing exponents in the operands is a key consideration. Direct addition is not feasible due to this potential discrepancy in exponents. To address this, our approach involves scaling the operand with the larger exponent to align with the smaller exponent, followed by addition, normalization, and rounding. Assuming, without loss of generality, that $e_a \geq e_b$, we initially scale $s_a$ to $s_a \cdot 2^\lambda$, where $\lambda = e_a - e_b$. This alignment enables us to compute the sum in a finite field as $\hat{s_c} = s_a \cdot 2^\lambda + s_b$. Following the computation of $\hat{s_c}$, normalization is applied in a manner similar to the processes for multiplication and division. Here, the exponent relationship is given by $e_b + \theta = e_c$. Consequently, the constraints can be formulated as follows:
\begin{equation}\label{eq:our_add}
  \begin{split}  
    s_a \cdot 2^\lambda + s_b - s_c \cdot 2^\theta &\leq \delta (s_a \cdot 2^\lambda + s_b),  \\
    e_c - \theta &= e_b, \\
    e_b + \lambda &= e_a
  \end{split}
\end{equation} 

There is one more thing to note: in the actual circuit, we do not know which of the two operands has the larger exponent. To manage this and ensure that $e_a \geq e_b$, we adopt a technique from \cite{CCS:ZFZS20}, which involves the use of a permutation gadget. The permutation gadget operates by rearranging the operands such that the exponent of the first operand in the addition process is guaranteed to be no less than that of the second operand. We give the detail of this gadget in Appendix \ref{app:A}.

\brief{Our generic circuits.} Now we discuss how to present the above constraints in a circuit. Our approach involves the prover initially performing the arithmetic as outlined and obtaining results along with some auxiliary inputs. The prover then demonstrates that these results and auxiliary inputs fulfill the requirements of the specified circuit.

Firstly, to see the $\leq$ relation in the above constraints holds, we utilize a \textsf{compare} gadget leveraging a generic bit-decomposition technique as \cite{SP:BBBPWM18} did. To prove that $x \leq y$, where $x$ and $y$ are $w$-bit integers, we compute $m = y - x + 2^{w}$ and subsequently decompose $m$ into $w+1$ bits. If the most significant bit of $m$ is 1, it indicates that $x \leq y$ is valid. Conversely, if $m_w$ is 0, then $x > y$. We leave the details of this comparison gadget in Appendix \ref{app:A}.

Addressing the remaining constraints in relations \ref{eq:our_mul}, \ref{eq:our_div}, and \ref{eq:our_add} requires handling the exponential computations involving $(\theta, 2^\theta)$ and $(\lambda, 2^{\lambda})$. This task presents two main challenges: firstly, the variables $\theta$ and $\lambda$ are non-deterministic and provided by the prover, necessitating their restriction within the circuit. Secondly, it takes a lot of gates for restricting the exponential relation. To optimize the circuit design for these constraints, we draw upon insights similar to those in \cite{CCS:GJJZ22}. First we consider the following lemma:

\begin{lemma}\label{lem:our_mul}
  In relation \ref{eq:our_mul}, if $s_a, s_b, s_c$ are all $w$-bits integers, then $\theta \in \{w-1, w\}$ always holds.
\end{lemma}

The same insights apply for Lemma \ref{lem:our_div}, \ref{lem:our_add} as well. We leave the proofs in Appendix \ref{app:B}. Leveraging these insights allows us to directly enforce the constants $w, w-1$, and their corresponding exponential values $2^w, 2^{w-1}$ into the circuit. We then constrain $\theta$ using a selector, which efficiently bounds its value and reduces the overhead associated with exponential relations. By integrating these elements, we formulate the final circuit \textsf{FloatMul} for the multiplication gate based on relation \ref{eq:our_mul}.
\begin{tcolorbox}[float, width=\columnwidth, boxrule=0.5pt,colback=white]
  $\mathsf{FloatMul}_w(a:(s_a, e_a), b:(s_b, e_b), c:(s_c, e_c))$: 
  \raggedright
  \begin{flalign*}
    1. & e_c - \theta := e_a + e_b \\
    2. & (\theta - w)(\theta - (w-1)) := 0 \\
    3. & \textsf{mid} := (\theta - (w - 1)) \cdot 2^w - (\theta - w) \cdot 2^{w-1} \\ 
    4. & x := s_a \cdot s_b, \quad y := s_c \cdot \mathsf{mid}, \quad z := \delta^{-1} \cdot (x - y) \\
    5. & 1 := \mathsf{compare}_{2w}(z, x) \\
    6. & 1 := \mathsf{bit\_decompose}_w(s_c, \{\mathbf{s_c}_i\}), \quad 1 := \mathbf{s_c}_0 &
  \end{flalign*}
  \noindent \rule[0.25\baselineskip]{\textwidth}{0.5pt}
  \textsf{mid} denotes for $2^\theta$. $\{\mathbf{s_c}_i\}$ denotes for the bit-decomposition of $s_c$. In the step 3 we restrict the value of \textsf{mid} using interpolation. The \textsf{compare} and \textsf{bit\_decompose} gadgets each takes $O(w)$ gates and the whole circuit contains approximately $3w$ constraints.
\end{tcolorbox}

The construction of gadgets used is detailed in Appendix \ref{app:A}. Regarding division operations, the circuit can be derived similarly.

\begin{lemma}\label{lem:our_div}
  In relation \ref{eq:our_div}, if $s_a, s_b, s_c$ are all $w$-bits integers, then $\theta \in \{w-1, w\}$ always holds.
\end{lemma}

\begin{tcolorbox}[float,width=\columnwidth, boxrule=0.5pt,colback=white]
  $\mathsf{FloatDiv}_w(a:(s_a, e_a), b:(s_b, e_b), c:(s_c, e_c))$:
  \raggedright
  \begin{flalign*}
    1. & e_c + \theta := e_a - e_b \\
    2. & (\theta - w)(\theta - (w-1)) := 0 \\
    3. & \textsf{mid} := (\theta - (w - 1)) \cdot 2^w - (\theta - w) \cdot 2^{w-1} \\
    4. & x := s_a \cdot \mathsf{mid}, \quad y := s_c \cdot s_b, \quad z := \delta^{-1} \cdot (x - y) \\
    5. & 1 := \mathsf{compare}_{2w}(z, x) \\
    6. & 1 := \mathsf{bit\_decompose}_w(s_c, \{\mathbf{s_c}_i\}), \quad 1 := \mathbf{s_c}_0 &
  \end{flalign*}
  \noindent \rule[0.25\baselineskip]{\textwidth}{0.5pt}
  The whole circuit contains approximately $3w$ constraints.
\end{tcolorbox}

For addition gates, the process begins with the derivation of Lemma \ref{lem:our_add}, which enables the enforcement of the relationship between $2^\lambda$ and $2^\theta$. However, a notable challenge arises with the computation of $(\lambda, 2^\lambda)$. Due to the potential for arbitrarily large gaps between the two exponents, directly enforcing $\lambda$ into the circuit is not feasible. To overcome this, we adopt a repeated-squaring method for computing $2^\lambda$ based on the given exponent difference $\lambda = e_a - e_b$. This approach can compute $2^\lambda$ using $O(\log{\lambda})$ gates, thus optimizing the circuit's efficiency. Putting things together, we obtain the addition circuit. We leave the detail of gadgets used in Appendix \ref{app:A} along with some further optimizations.

\begin{lemma}\label{lem:our_add}
  In relation \ref{eq:our_add}, if $s_a, s_b, s_c$ are all $w$-bits integers, then $\theta \in \{\lambda, \lambda+1\}$ always holds.
\end{lemma}

\begin{tcolorbox}[float,width=\columnwidth, boxrule=0.5pt,colback=white]
  $\mathsf{FloatAdd}_w(a:(s_a, e_a), b:(s_b, e_b), c:(s_c, e_c))$:
  \raggedright
  \begin{flalign*}
    1. & 1 := \mathsf{permutation\_check}((e_a, s_a), (e_b, s_b), (\hat{e_a}, \hat{s_a}), (\hat{e_b}, \hat{s_b})) \\
    2. & 1 := \mathsf{compare}(\hat{e_b}, \hat{e_a}) \\
    3. & \hat{e_b} = e_c - \theta, \quad \hat{e_a} - \hat{e_b} = \lambda \\
    4. & (\theta - \lambda)(\theta - (\lambda+1)) = 0 \\
    5. & 1 := \mathsf{exponential\_check}(\lambda, \mathsf{mid'}) \\
    6. & \mathsf{mid} := \mathsf{mid'} \cdot (\theta - \lambda + 1) \\
    7. & x = \hat{s_a} \cdot 2^\lambda + \hat{s_b}, \quad y = s_c \cdot \mathsf{mid}, \quad z = \delta^{-1} \cdot (x - y) \\
    8. & 1 := \mathsf{compare}_{2w}(z, x) \\
    9. & 1 := \mathsf{bit\_decompose}_w(s_c, \{\mathbf{s_c}_i\}), \quad 1 := \mathbf{s_c}_0 &
  \end{flalign*}
  \noindent \rule[0.25\baselineskip]{\textwidth}{0.5pt}
  $\mathsf{mid'}$ denotes for $2^\lambda$, $(\hat{e_a}, \hat{s_a}), (\hat{e_b}, \hat{s_b})$ denotes for the permuted value of $(e_a, s_a), (e_b, s_b)$ that satisfies $\hat{e_b} \leq \hat{e_a}$. In step 6 we restrict the relation between $2^\lambda$ and $2^\theta$ using an interpolation. The \textsf{exponential\_check} and \textsf{permutation\_check} gadgets take $O(\log\lambda)$ and $O(1)$ gates, respectively. The whole circuit contains approximately $3w+3\log\lambda$ constraints.
\end{tcolorbox}

\brief{Analysis.} For security, since circuits for ZKP systems are generally designed on a finite field with a specified size, the computation in the circuits should not "wrap-around" (meaning values exceed the finite field). To make the relations \ref{eq:our_mul}, \ref{eq:our_div}, \ref{eq:our_add} hold on the field (the computation will not wrap-around), we need the prime of the field satisfies $p > 2^{3w+1}$. We defer the proof of this claim to Appendix \ref{app:C}.

For efficiency, our generic circuits take $O(w)$ gates for a single decimal arithmetic operation. We implement the above circuits for 32-bits floating-point numbers ($w$=23). It takes 131 gates to implement an addition gate in a circuit. Both the multiplication circuit and the division circuit take 82 gates. Compared with \cite{USENIX:WYXKW21}, It improves the efficiency nearly $75 \times$. We note that in \cite{CCS:GJJZ22}, the authors get a similar result (Estimated roughly requiring 108 gates for addition and 25 gates for multiplication under a finite field where $p=2^{256}$). However, their method introduces additional $O(w^2)$ overhead outside the circuit, thus cannot be conveniently integrated with an existing ZKP backend. And they haven't provided a concrete implementation either. Thus, our result can be considered as the first implementation of generic circuits for decimal arithmetic that can be seamlessly integrated with an existing ZKP backend.

In the actual implementation, we set $w=23$ and $\delta = 2^{-22}$. But the precision can be adjusted according to the actual needs.

\subsection{Put everything together} \label{sec:zkTI-construction}

With the specialized circuits in place, we can now construct the circuit for the truth inference algorithm $f$. For the algorithms selected, we transform them into circuits by substituting each instance of decimal arithmetic (i.e., addition, multiplication, and division) and other operations with the corresponding circuits we have designed.

In our system, built upon a generic crowdsourcing framework, the aggregator \aggregator, after receiving responses $\answer$ from workers, executes one iteration of the truth inference algorithm $f$ (as selected in section \ref{sec:algos}) to deduce the inferred truth and the updated quality of workers, denoted as $\answer^*, \quality = f(\answer, \prior)$. \aggregator also obtains some auxiliary inputs during the decimal arithmetic computation. To validate these results, entities initially prepare commitments as outlined in section \ref{sec:po}. Subsequently, during the \textsf{Prove} phase, \aggregator transforms $f$ and the process of opening commitments into a circuit \circuit. Utilizing an existing ZKP backend, \aggregator then generates a proof $\pi$. This proof is ultimately subject to a verifier.

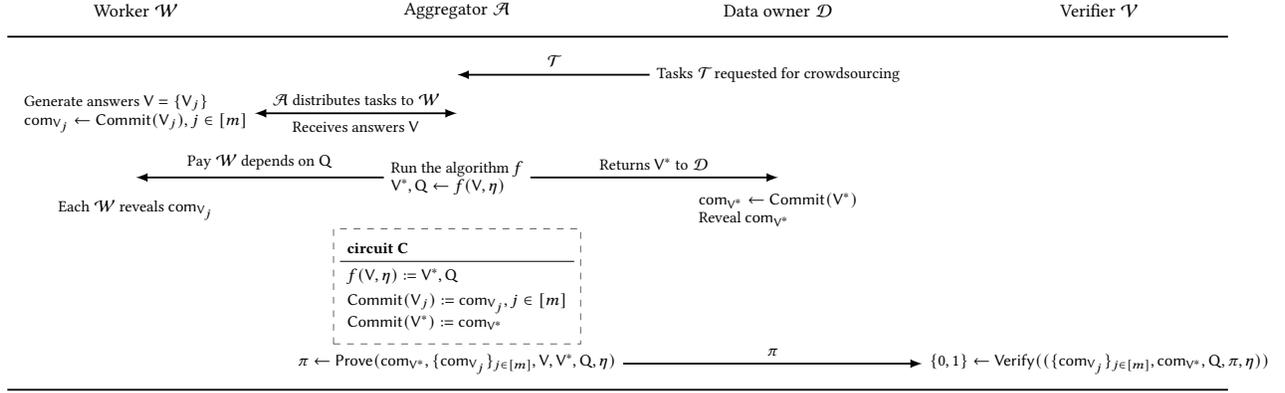
\begin{figure*}[!h] 
  \centering
  \resizebox{2\columnwidth}{!}{%
    \begin{tikzpicture}
      \tikzstyle{line} = [thick,>=latex];
      \tikzstyle{comment} = [text centered, font=\footnotesize];
      \tikzstyle{party} = [rounded corners, inner sep=2mm, minimum height=.8cm,rectangle, font=\small];
      \tikzstyle{circuit} = [rectangle, draw=black!60, dashed, inner sep=1mm];
      \node [party] (worker) at (1,0) {Worker \worker};
      \node [party] (aggregator) at (6,0) {Aggregator \aggregator};
      \node [party] (data-owner) at (11,0) {Data owner \dataowner};
      \node [party] (verifier) at (16,0) {Verifier \verifier};

      \draw [line] ([xshift=-2cm] worker.south) -- ([xshift=2cm] verifier.south);

      \scoped[yshift=-1cm]{
        \node [comment] (data-owner-tasks) at (11,0) {Tasks \task requested for crowdsourcing};
        \draw [->, line] (data-owner-tasks.west) -- node[midway,above,comment]{\task} ([yshift=-1cm]aggregator.center);
      }%
      \scoped[yshift=-1.6cm]{
        \node [comment, align=left] (worker-actions) at (1,0) {Generate answers $\answer = \{\answer_j\}$ \\ $\mathsf{com}_{\answer_j} \leftarrow \mathsf{Commit}(\answer_j), j\in[m]$};
        \draw [<->, line] (worker-actions.east) -- node[midway,above,comment]{\aggregator distributes tasks to \worker} node[midway,below,comment]{Receives answers \answer} (6,0);
      }%
      \scoped[yshift=-2.6cm]{
        \node [comment, align=left] (aggregation) at (6,0) {Run the algorithm $f$ \\ $\answer^*, \quality \leftarrow f(\answer, \prior)$};
        \draw [->, line] (aggregation.east) -- node[midway,above,comment]{Returns $\answer^*$ to \dataowner} (11,0);
        \draw [->, line] (aggregation.west) -- node[midway,above,comment]{Pay \worker depends on \quality} (1,0);
      }%
      \scoped[yshift=-3.1cm]{
        \node [comment, align=left] (reveal-comA) at (1,0) {Each \worker reveals $\mathsf{com}_{\answer_j}$};
        \node [comment, align=left] (reveal-comV) at (11,0) {$\mathsf{com}_{\answer^*} \leftarrow \mathsf{Commit}(\answer^*)$ \\ Reveal $\mathsf{com}_{\answer^*}$};
      }%
      \scoped[yshift=-4.5cm]{
        \node [comment, align=left] (circuits) at (6,0) {$f(\answer, \prior) \assign \answer^*, \quality$ \\[0.1cm] $\mathsf{Commit}(\answer_j) \assign \mathsf{com}_{\answer_j}, j\in[m]$ \\ $\mathsf{Commit}(\answer^*) \assign \mathsf{com}_{\answer^*}$};
        \node [comment, above=6pt of circuits.north west, anchor=west] (circuits-header) {\textbf{circuit \circuit}};
        \draw (circuits.north west) -- (circuits.north east);
        \node [circuit, fit=(circuits) (circuits-header)] (circuits-container){};
        \node [comment, align=left] (prove) at (6, -1){$\pi \leftarrow \mathsf{Prove}(\mathsf{com}_{\answer^*}, \{\mathsf{com}_{\answer_j}\}_{j\in[m]}, \answer, \answer^*, \quality, \prior)$};
      
        \node [comment, align=left] (verify) at (16, -1) {$\{0,1\} \leftarrow \mathsf{Verify}((\{\mathsf{com}_{\answer_j}\}_{j\in[m]}, \mathsf{com}_{\answer^*}, \quality, \pi, \prior))$};
        \draw [<-, line] (verify.west) -- node[midway,above,comment]{$\pi$} (prove.east);
      }%

      \draw [line] ([xshift=-2cm, yshift=-5.5cm] worker.south) -- ([xshift=2cm, yshift=-5.5cm] verifier.south); 
    \end{tikzpicture}
  }%
  \caption{The complete workflow of the \zkTI protocol in our system. Upon the foundation of the crowdsourcing process, \aggregator proves the correctness of the algorithm and the consistency of the inputs, allowing any entity to verify it.}
  \label{fig:full-protocol}
\end{figure*}

We synthesize all components and formally delineate our construction in Protocol \ref{protocol:1}. We also provide a comprehensive depiction of our system's workflow in Figure \ref{fig:full-protocol}. Furthermore, we have the following theorem:

\begin{theorem}\label{thm:1}
   Protocol \ref{protocol:1} is a zero-knowledge truth inference protocol by Definition \ref{def:zkTI}.
\end{theorem}

We leave the security proof of Theorem \ref{thm:1} in Appendix \ref{app:C}.

\begin{protocolbox}[protocol:1]
  \footnotesize\textsc{Protocol 1 (zero-knowledge Truth Inference):} \\
  \small\textit{
    Let $\lambda$ be a security parameter, $\FF$ be a finite field, $\answer$ be the answers from each data sources, $f$ is a truth inference algorithm. \circuit is the overall circuit compiles $f$ and the opening of all commitments. \aggregator is the prover and \verifier is a verifier. $\mathsf{ZKP}$ is the underlying ZKP system. \aux represents the auxiliary inputs (extended witness) used for proving the decimal arithmetic circuits.
  }
  \begin{itemize}[nosep, leftmargin=0.3cm, itemindent=0cm]
    \raggedright
    \item $\mathsf{pp} \leftarrow \mathsf{zkTI.Setup(1^\lambda)}$: let $\mathsf{pp_1} = \mathsf{ZKP.Setup}(1^\lambda)$, \prior to be priority factors of $f$, $\mathsf{pp} = \{\mathsf{pp}_1, \prior\}$.
    \item $\{\mathsf{com}_{\answer_j}\}_{j\in[m]}, \mathsf{com}_{\answer^*} \leftarrow \mathsf{zkTI.Commit}(\{\answer_j\}_{j\in[m]}, \answer^*)$: 
    Each worker of \worker commits to $\answer_j$ as $\mathsf{com}_{\answer_j} = \mathsf{Commit}(\answer_j)$. \aggregator commits to $\answer^*$ as $\mathsf{com}_{\answer^*} = \mathsf{Commit}(\answer^*)$.
    \item $\pi \leftarrow \mathsf{zkTI.Prove}(\{\mathsf{com}_{\answer_j}\}_{j\in[m]}, \mathsf{com}_{\answer^*}, \answer, \answer^*, \quality, \aux, \mathsf{pp})$: \aggregator received the answers \answer and run the $\answer^*, \quality = f(\answer, \prior)$ to get the inferred result along with the qualities of each worker. He additionally obtains \aux from the computation process. Let $w = \{\answer, \answer^*, \aux\}$ and $x =\{ \{\mathsf{com}_{\answer_j}\}_{j\in[m]}, \mathsf{com}_{\answer^*}, \quality, \prior \}$. \aggregator invoke $\mathsf{ZKP.Prove}(\circuit, x, w, \mathsf{pp_1})$  based on \circuit to get the proof $\pi$. \aggregator sends $\pi$ to \verifier.
    \item $\{\mathsf{0,1}\} \leftarrow \mathsf{zkTI.Verify}(x, \pi, \mathsf{pp_1})$: \verifier accepts $\pi$ if $\mathsf{ZKP.Verify}(\circuit, x, \pi, \mathsf{pp_1})$ output 1, otherwise rejects.
  \end{itemize}
\end{protocolbox}


%% file: sections/evaluation.tex

We fully implemented our ideas. In this section we introduce our implementation details and evaluate the performance of our protocol.

\subsection{Implementation}

Our implementation process commences with the realization of the protocol as delineated in section \ref{sec:fpc}. Following the guidance provided, we developed circuits for decimal arithmetic. Utilizing the open-source zero-knowledge proof framework \textsc{Libsnark}~\cite{noauthor_libSNARK_nodate}, we successfully transformed the algorithms \CRH and \ZenCrowd into circuit forms. Additionally, for comparative analysis, we implemented a straightforward \MV algorithm, which is detailed in section \ref{sec:setup}. The \textsc{Libsnark} framework supports a 254-bit field, meeting our security requirements for decimal arithmetic circuits. Consequently, we completed the full protocol implementation.

\brief{Choice of the zero-knowledge proofs backend.} We choose \textsc{Groth16}~\cite{EC:Groth16} and \textsc{Spartan}~\cite{C:Setty20} as our backends for evaluation. \textsc{Groth16} is a pairing-based zkSNARK system characterized by minimal verification overhead but a relatively high proving overhead. Conversely, \textsc{Spartan} excels in proof generation efficiency but has sublinear verification overhead and proof size. These attributes enable us to optimize efficiency by choosing appropriate backends for different scenarios. While \textsc{Libsnark} inherently supports \textsc{Groth16}, it lacks native support for \textsc{Spartan}. To facilitate the latter, we developed a pipeline capable of exporting gates and variables from \textsc{Libsnark} and importing them into \textsc{Spartan}, conforming to the \textsc{Zkinterface} binary-file standard~\cite{noauthor_zkInterface_nodate}.

\brief{Collision-resistant hashing function.} As mentioned in section \ref{sec:po}, the commitment scheme is underpinned by a hashing function. Our criteria for selecting a hashing function revolved around its circuit efficiency. We settled on \textsc{Poseidon}~\cite{USENIX:GKRRS21} due to its permutation-based nature that induces relatively low circuit overhead, and its provision of a 128-bit security level within a 254-bit field, aligning well with our application's security needs. Implementing \textsc{Poseidon} in the arity-4 setting, it requires approximately 16K gates to hash $n=100$ elements into a single commitment. For $m=30$ workers, the total commitment cost approximates to 460K gates.

In summary, our development involved over 3000 lines of C++ code for the protocol and pipeline implementation. Additionally, we adapted an open-source library~\cite{noauthor_spartaninterface_nodate} with about 300 lines of Rust code to facilitate \textsc{Zkinterface} file integration into \textsc{Spartan}. The truth inference algorithms \CRH and \ZenCrowd were also implemented in roughly 1000 lines of Python code for accuracy comparison. Our codebase is openly accessible at \url{https://anonymous.4open.science/r/zkTI}.

\subsection{Experimental setup} \label{sec:setup}

\brief{Baseline.} In \cite{ASIACCS:XLXRZSD20}, the authors developed a system named \textsc{V-patd}, utilizing \textit{bilinear-pairing} and \textit{signature} techniques for verifiable computing in crowdsourcing truth inference. This system was applied to the \CRH algorithm~\cite{li_resolving_2014}, making it verifiable. Due to the similarity in goals and the algorithm employed, we have chosen this work as a baseline for our comparative analysis.

\brief{Hardware.} Since the implementation from \cite{ASIACCS:XLXRZSD20} was not open-sourced, we opted for comparable hardware specifications to ensure a fair comparison. Our experiments were conducted on a server equipped with 16 GB of RAM and a 2.50GHz Intel(R) Xeon(R) Platinum 8269CY CPU.

\brief{Dataset.} For the evaluation, we primarily focused on decision-making tasks, where algorithms infer truths based on binary (0/1) responses from workers. The simplistic \MV algorithm determines truth via majority voting, yet it is susceptible to manipulation by malicious workers. In contrast, both \CRH and \ZenCrowd incorporate worker quality into their calculations, enhancing resilience against such attacks. It's worth noting that our protocol also supports other types of tasks, as demonstrated in \cite{zheng_truth_2017}.

Our primary concern is the efficiency of our protocol, specifically the overhead incurred and how it scales with varying numbers of tasks and workers. The content of tasks and the accuracy of algorithms are secondary considerations. For those interested in algorithmic accuracy, further details can be found in \cite{zheng_truth_2017}. In our experiments, we utilized two types of datasets. Synthetic datasets, generated through random binary answers, were used to assess the efficiency and scalability of our protocol, with variable numbers of tasks and workers. For practical applicability, we employed a real-world dataset from \cite{zheng_truth_2017}, encompassing 108 tasks, responses from 39 workers, and a total of 4212 answers. The objective is to infer truths for each task and assess the quality of each worker. The initial quality of workers was set at 0.5, within a range of [0, 1].

\subsection{Evaluation}

In this section, we present the evaluation results of our \zkTI protocol. Initially, experiments were conducted on a real-world dataset, where the \CRH and \ZenCrowd algorithms achieved an accuracy of 78.4\%, surpassing the 75.9\% accuracy of the \MV algorithm. This consistency with the results obtained from plain Python code execution validates our design's correctness and applicability to real-world scenarios. Subsequently, our focus shifted to assessing the protocol's efficiency, encompassing the computational overhead for both the aggregator (prover) and the verifier.

\subsubsection{Comparison with baseline work}

\begin{figure}[h]
  \centering
  \begin{subfigure}[b]{0.47\linewidth}
      \includegraphics[width=\textwidth]{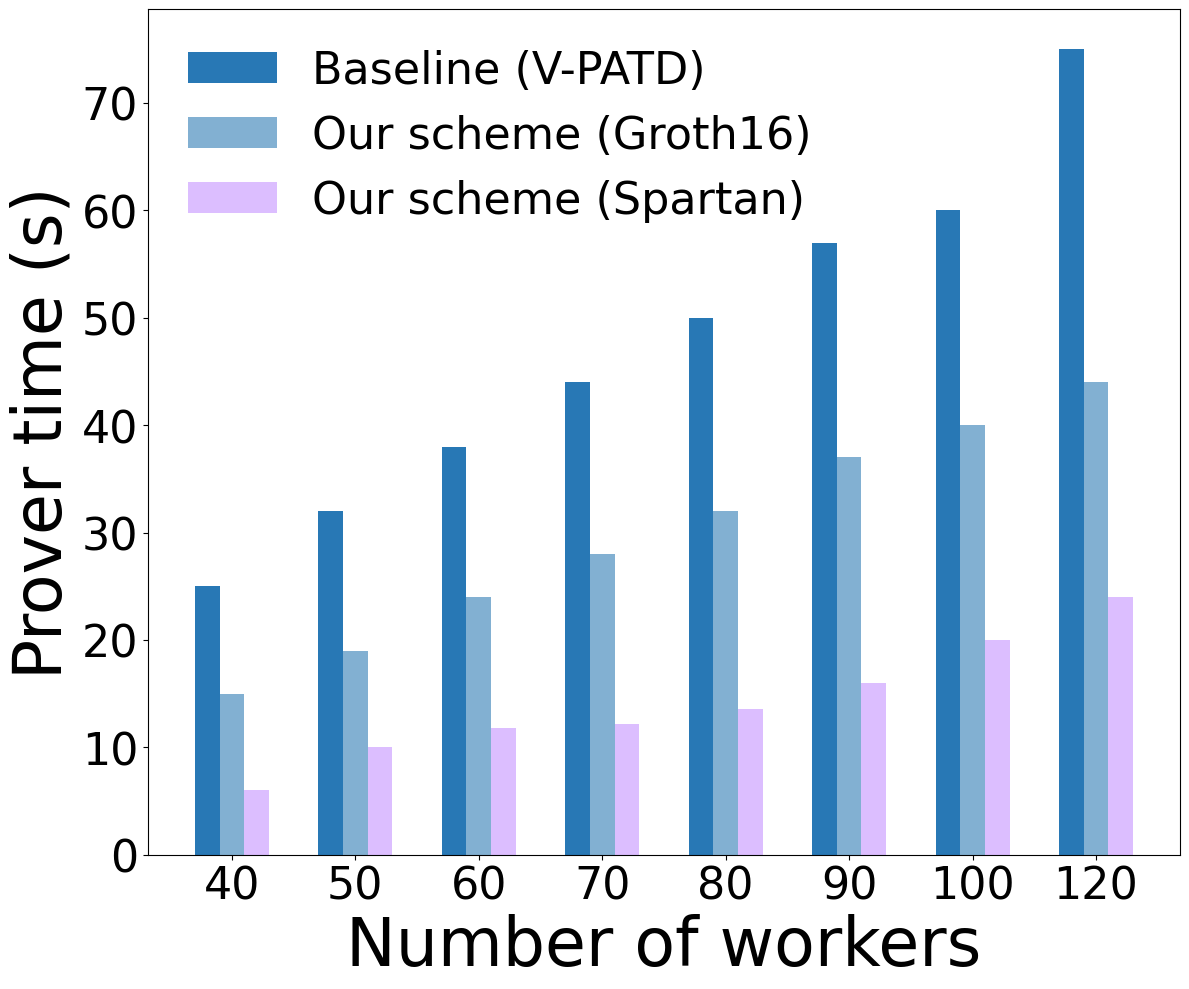}
      \caption{}
      \label{fig:fig2-a}
  \end{subfigure}
  \hspace{0.25cm} 
  \begin{subfigure}[b]{0.47\linewidth}
      \includegraphics[width=\textwidth]{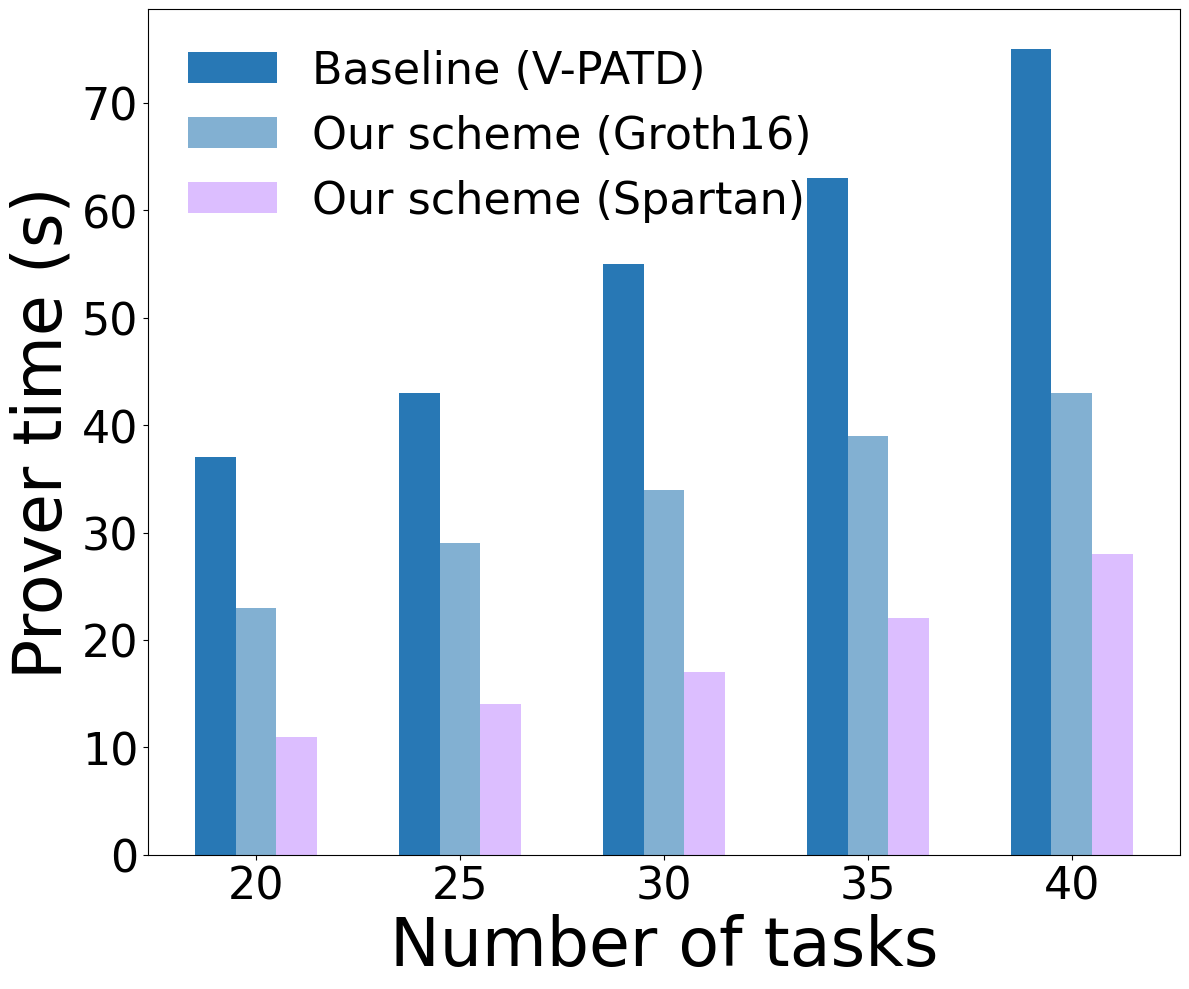}
      \caption{}
      \label{fig:fig2-b}
  \end{subfigure}
  \caption{Prover time (aggregator overhead) comparison with \cite{ASIACCS:XLXRZSD20} applying \textsf{CRH}. (a) $n=25$, with the different number of workers. (b) $m=75$, with the different number of tasks.}
  \label{fig:fig2}
\end{figure}

We benchmarked our protocol against the baseline work \cite{ASIACCS:XLXRZSD20} using the \CRH algorithm. In our implementation, \CRH, along with the opening of commitments, was formulated into a circuit. The experiments varied either the number of workers or tasks while keeping the other constant. These circuits were then integrated into the \textsc{Groth16} and \textsc{Spartan} systems for analysis. The primary bottleneck, the aggregator's computational overhead for proof generation, is depicted in Figure \ref{fig:fig2}.

The experiment with \CRH used datasets with decimal arithmetic precision set at $w=23$. Two key observations emerged: firstly, our protocol significantly reduced overhead compared to the baseline, showing a 1.5 to 4 times improvement depending on backend and parameters. Secondly, the computational overhead correlates with the product of the number of tasks and workers, aligning with the algorithm's need to traverse all tasks and workers for calculations. We also report that for verifiers, \textsc{Groth16} incurs minimal verification overhead, while \textsc{Spartan} adds a verification time of 1-2.5 seconds and a communication overhead (proof size) of $0.2$MB$-$$0.6$MB, which remains acceptable for most applications.

\subsubsection{Performance with different dataset size}

Subsequently, we assessed our protocol's efficiency under various settings, measuring the computational overhead for both the aggregator (prover) and the verifier, as well as the communication costs (proof size). Initially, we fixed the dataset size ($|\worker| \times |\task|$) at $100 \times 30$, testing different algorithms across backends. Performance metrics under these settings are tabulated in Table \ref{tab:2}, which also includes the performance of the \MV algorithm for comparison. Despite its smaller circuit size, the \MV algorithm remains susceptible to attacks by malicious workers.

\begin{figure}[h]
  \centering
  \includegraphics[width=0.48\textwidth]{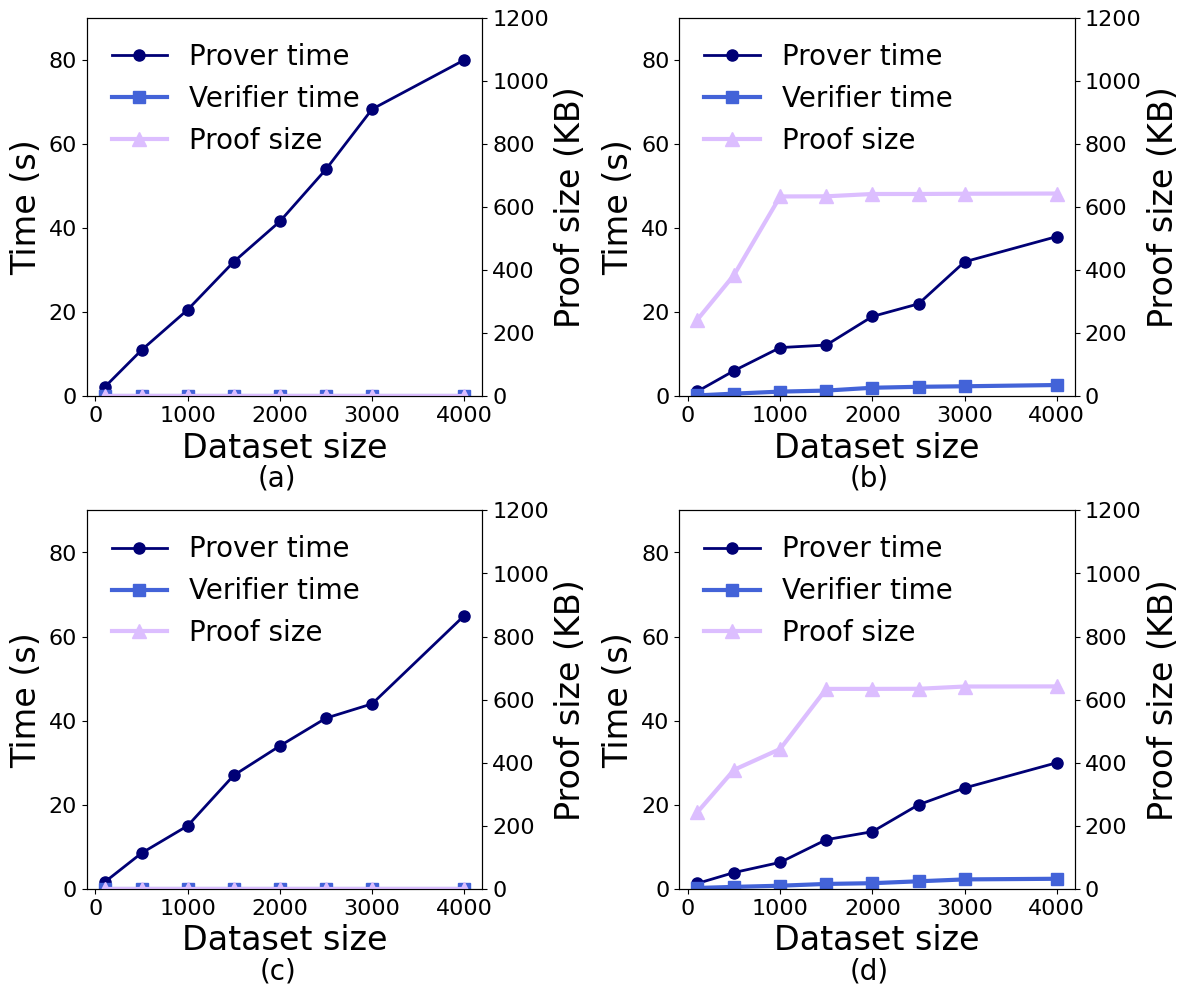}
  \caption{Performance of algorithms under different backends varying dataset size. (a) \textsc{Groth16} for \textsf{ZC} (b) \textsc{Spartan} for \textsf{ZC} (c) \textsc{Groth16} for \textsf{CRH} (d) \textsc{Spartan} for \textsf{CRH}. In each subfigure we measure the prover time (aggregator overhead), verifier time (verification overhead) and proof size (communication overhead). In (a) (c), the verifier time and the proof size are tiny, both laying at the bottom of the subfigure.}
  \label{fig:fig3}
\end{figure}

Lastly, we varied the dataset size from 100 to 4000, benching the performance of different algorithms under each backend, and presented the results in Figure \ref{fig:fig3}. Notably, by comparing the subfigures in the same column, we can observe that \ZenCrowd has a higher computational overhead than \CRH with the same backend and dataset size. As the dataset size increases, the runtime for each backend correspondingly rises. \textsc{Spartan} maintains the fastest proof generation speed, generating proofs for datasets as large as 4000 in under 40 seconds for both algorithms. Conversely, \textsc{Groth16}'s significant memory consumption could lead to failure in larger circuits, while \textsc{Spartan} is expected to handle larger scales more effectively. Through the comparison of subfigures in the same row, we obtain that \textsc{Groth16}'s advantage lies in its minimal verification overhead, making it suitable for specific scenarios like blockchain oracles. However, with \textsc{Spartan}, verification and communication overheads increase proportionally. Overall, both backends demonstrate enhanced efficiency compared to \cite{ASIACCS:XLXRZSD20}, underscoring the scalability of our protocol.

\begin{table}[t]
  \centering
  \begin{tabular}{cccccc} 
    \toprule 
    \multirow{2}*{} & \multirow{2}*{types} & \multirow{2}*{$|\mathsf{C}|$}  & \multicolumn{2}{c}{time budget} & \multirow{2}*{$|\pi|$} \\
    \cline{4-5}
    \addlinespace
    & & & \textsf{P} & \textsf{V} & \\ 
    \midrule 
    \addlinespace
    \multirow{2}{*}{\textsf{MV}} & \small\textsc{Groth16} & \multirow{2}{*}{0.57M} & 3.98s & 1ms & 1KB \\
                                & \small\textsc{Spartan} &  & 2.0s & 0.36s & 301KB \\
                                \addlinespace
    \multirow{2}{*}{\textsf{ZC}} & \small\textsc{Groth16} & \multirow{2}{*}{2.21M} & 68.3s & 1ms & 1KB \\
                                & \small\textsc{Spartan} &  & 32s & 2.4s & 658KB \\
                                \addlinespace
    \multirow{3}{*}{\textsf{CRH}} & \small\textsc{Groth16} & \multirow{2}{*}{1.76M} & 44.9s & 1ms & 1KB \\
                                  & \small\textsc{Spartan} &  & 24s & 2.31s & 657KB \\
                                  & \small\textsc{V-patd} & --- & 75.0s$^*$ & --- & --- \\
    \bottomrule 
  \end{tabular} 
  \vspace{0.5cm}
  \caption{
    Performance of algorithms under different backends (with dataset size of $100 \times 30$). $|\mathsf{C}|$ denotes for circuit size, containing the costs of both the algorithm applied and the opening of commitments. $|\pi|$ denotes for proof size. $*$ means the value is estimated. 
  } \label{tab:2}
\end{table}

\subsubsection{Decimal arithmetic circuits with different precision}

In our final analysis, we explored the trade-off between circuit size and decimal arithmetic precision by adjusting the precision setting $w$. This adjustment directly influences the gate requirements for each arithmetic operation. In Table \ref{tab:3}, we present the number of gates needed for each arithmetic operation under different precision modes. Notably, modes with $w=16$ and $w=8$ can reduce the circuit size to $78.7\%$ and $54.7\%$, respectively, of the original circuit size at $w=23$. This finding highlights the flexibility of our decimal arithmetic scheme.

By compromising on precision to a certain extent, our approach achieves a reduction in circuit size, which in turn enhances efficiency. The precision setting can be tailored to meet the specific demands of different applications. This adaptability allows for a balanced approach between achieving computational efficiency and maintaining adequate precision for the task at hand.

\begin{table}[t]
  \centering
  \begin{tabular}{cccc} 
    \toprule 
    {} & \textsf{FloatAdd} & \textsf{FloatMul}  & \textsf{FloatDiv} \\
    $w=23$ & 131 & 82 & 82 \\
    $w=16$ & 110 & 61 & 61 \\
    $w=8$ & 86 & 37 & 37 \\
    \bottomrule 
  \end{tabular} 
  \vspace{0.5cm}
  \caption{
    The gates needed for each decimal arithmetic operation with different $w$ setting. Here $w$ is the bit number of the significand number in floating-points.
  } \label{tab:3}
\end{table}

%% file: sections/conclusion.tex

In this work we propose a novel approach, replacing outsourcing with crowdsourcing, to address the trust issues in outsourced computation. We primarily offer two contributions: (i) present the \zkTI protocol, making crowdsourcing truth inference algorithms verifiable while considering workers' quality and (ii) new techniques for expressing decimal arithmetic in circuits. Our work can be applied in various scenarios such as data annotation, question answering systems, and blockchain oracles, thereby establishing bridges of trust. The techniques for decimals can be applied in other scenarios requiring high-precision computations, such as \textit{zkML} and \textit{DeFi}.

%% file: sections/appendix.tex

\section{Construction of gadgets} \label{app:A}

\begin{tcolorbox}[width=\columnwidth, boxrule=0.5pt,colback=white]
  $\mathsf{bit\_decompose}_w(v, \{v_i\})$:
  \raggedright
  \begin{flalign*}
    1. & v :=  \sum_{i=0}^{w-1}2^i \cdot v_i \\
    2. & (1-v_i) \cdot v_i = 0 \text{, \quad For each } i \in [0,w)& 
  \end{flalign*}
  \noindent \rule[0.25\baselineskip]{\textwidth}{0.5pt}
  Constraints for checking $\{v_i\}$ is a bit-decomposition of $v$. Returns 1 if above constraints satisfies. The gadget takes $w+1$ gates.
\end{tcolorbox}

\begin{tcolorbox}[width=\columnwidth, boxrule=0.5pt,colback=white]
  $\mathsf{compare}_w(a, b)$:
  \raggedright
  \begin{flalign*}
    1. & m := b - a + 2^w \\
    2. & 1 := \mathsf{bit\_decompose}_{w+1}(m, \{m_i\}), \quad 1 := m_0 & 
  \end{flalign*}
  \noindent \rule[0.25\baselineskip]{\textwidth}{0.5pt}
  Constraints for $a \leq b$. Returns 1 if above constraints satisfies. The gadget takes approximately $w+2$ multiplication gates.
\end{tcolorbox}

\begin{tcolorbox}[width=\columnwidth, boxrule=0.5pt,colback=white]
  $\mathsf{exponential\_check}(a, b)$:
  \raggedright
  \begin{flalign*}
    1. & 1 := \mathsf{bit\_decompose}_{\log{a}}(a, \{a_i\}) \\ 
    2. & b_{i+1} := b_i \cdot (1 - a_i) + b_i \cdot 2^{2^i} \cdot a_i \text{, for each } i \in [0, \log{a}] \\
    3. & b_{\log{a}} := b, \quad b_0 := 1 &
  \end{flalign*}
  \noindent \rule[0.25\baselineskip]{\textwidth}{0.5pt}
  Constraints for checking $b = 2^a$, suppose $a$ has $\log{w}$-bits. Returns 1 if above constraints satisfies. $\{2^{2^i}\}$ will be initialized as a public input thus doesn't have a cost. The gadget takes approximately $O(\log{a})$ gates.
\end{tcolorbox}

\begin{tcolorbox}[width=\columnwidth, boxrule=0.5pt,colback=white]
  $\mathsf{permutation\_check}(\{(a_1, b_1), (a_2, b_2)\}, \{(\hat{a_1}, \hat{b_1}), (\hat{a_2}, \hat{b_2})\})$:
  \raggedright
  \begin{flalign*}
    1. & c_1 = a_1 + z \cdot b_1, c_2 = a_2 + z \cdot b_2 \\ 
    2. & c_3 = \hat{a_1} + z \cdot \hat{b_1}, c_4 = \hat{a_2} + z \cdot \hat{b_2} \\
    3. & (r - c_1)(r - c_2) = (r - c_3)(r - c_4) &
  \end{flalign*}
  \noindent \rule[0.25\baselineskip]{\textwidth}{0.5pt}
  Constraints for proving $\{(a_1, b_1), (a_2, b_2)\}$ is a pairwise permutation of $\{(\hat{a_1}, \hat{b_1}), (\hat{a_2}, \hat{b_2})\}$. Return 1 if above constraints satisfies. $r, z$ are random points provided by the verifier. Refer to \cite{CCS:ZFZS20} for the completeness and soundness of this construction. The gadget takes $O(1)$ gates.
\end{tcolorbox}

In section \ref{sec:fpc}, we give a construction of \textsf{FloatAdd} with the overhead of approximately $3w+3\log\lambda$ where $\lambda = e_a - e_b$ (suppose $e_a > e_b$). The $\log\lambda$ overhead comes from the exponential check for $\mathsf{mid}'= 2^\lambda$, which needs $O(\log\lambda)$ gates. 

In the actual implementation, to further reduce the overhead of the repeating-square method, $\lambda$ can be restricted to a certain range. That is, if $e_a - e_b$ exceeds this range, we let $c = \max{(a,b)}$ directly. This makes sense because a larger $\lambda$ means the difference between the two exponents is getting bigger. When the difference gets too large, the smaller operand can be ignored directly. Moreover, it can be shown that when $\lambda$ exceeds a certain threshold $\epsilon$, the accuracy of the addition operation still aligns with our precision requirements. For instance, if the desired precision is $\delta = 2^{-(w-1)}$, we can directly use $c = \max{(a,b)}$ whenever $e_a - e_b > \epsilon = w$. This adjustment effectively reduces the computational overhead from $O(\log\lambda)$ to $O(\log w)$, particularly when $\lambda$ is relatively large. If we set $(s_c, e_c) = \max(a, b) = (s_a, e_a)$ directly, it should satisfy that:
\begin{equation*}
  \begin{split}
    s_b \cdot 2^{e_b} \leq \delta (s_a \cdot 2^{e_a} + s_b \cdot 2^{e_b}) &\iff \\
    s_b \leq \delta (s_a \cdot 2^\lambda + s_b) &\iff \\
    \lambda \geq \log(\frac{s_b}{s_a}\cdot(\delta^{-1}-1)) 
  \end{split}
\end{equation*} 

Since $\log(\frac{s_b}{s_a}) \leq 1$, the relation is hold when $\lambda > 1 - \log\delta$. For example, if we require the precision $\delta = 2^{-(w-1)}$, then $\epsilon = w$. If we get a $\lambda_0 > w$, then we can output $c = \max(a,b)$ directly. Through this way, we decrease the overhead of addition from $O(\log\lambda)$ to $O(\log{w})$.

\section{Proof of Lemma \ref{lem:our_mul}, \ref{lem:our_div}, \ref{lem:our_add}} \label{app:B}

\begin{proof} 
  For lemma \ref{lem:our_mul}, since $s_a, s_b, s_c \in [2^{w-1}, 2^w)$ are $w$-bits integer numbers, then $s_a \cdot s_b = \hat{s_c} \in [2^{2w-2}, 2^{2w})$ holds (as it is an integer, more precisely, this range is $[2^{2w-2}, 2^{2w}-2^{w+1}+1]$). To normalize $s_c$ to $w$-bit again, we should right shift $\hat{s_c}$ by $\theta$ bits. Considering the case of upper and lower bounds, it necessitates to right shift $w-1$ bits for $2^{2w-2}$ and $w$ bits for $2^{2w}$ to falling back to the required interval. So for any value in this range, $\theta$ can take these two values. Oppositely, for any number to right shift $w+1$ bits, the result will fall in the range of $[2^{w-3}, 2^{w-1}-1]$, which contradicts our initial definition. For $w-2$ we can get the similar conclusion. In summary, we draw the conclusion that $\theta \in \{w-1, w\}$.

  For lemma \ref{lem:our_div}, the situation is similar.

  For lemma \ref{lem:our_add}, first let's recall the computation process. We have $s_a \cdot 2^\lambda + s_b = \hat{s_c}$. Next, we shift the $\hat{s_c}$ to the right, which is equivalent to dividing by $2^\theta$. We round the value to floor and get $s_c = \lfloor \frac{\hat{s_c}}{2^\theta} \rfloor$. To prove the lemma, first we can get that $\hat{s_c} \in [(2^\lambda+1)(2^{w-1}), (2^\lambda+1)(2^w-1)]$. Then, if we take $\theta = \lambda$ for the left bound, the result will fall in the range of $[2^{w-1}, 2^w)$, which is valid. To see this, we can derive like below:
  for $\geq 2^{w-1}$ holds, $\frac{(2^\lambda+1)(2^{w-1})}{2^\lambda} \geq 2^{w-1} $ which is straightforward.
  for $< 2^{w}$ holds,
  \begin{equation*}
    \begin{split}
      \frac{(2^\lambda+1)(2^{w-1})}{2^\lambda} < 2^{w} &\iff \\
     (2^\lambda+1)(2^{w-1}) < 2^{w} \cdot 2^\lambda &\iff \\
     2^{w+\lambda} - 2^{\lambda+w-1}-2^{w-1} > 0 &\iff \\
     2^\lambda > 1 
    \end{split}
  \end{equation*}
  which is hold. Then we consider for the right bound. This is more complex than before. To see this, if we take $\theta = \lambda+1$ for the right bound, to prove $ < 2^{w}$ holds, we derive from following:
  \begin{equation*}
    \begin{split}
      \frac{(2^\lambda+1)(2^{w}-1)}{2^{\lambda+1}} < 2^{w} &\iff \\
     (2^\lambda+1)(2^{w-1}) < 2^{w+\lambda+1} &\iff \\
     2^{w+\lambda} + 2^{\lambda}- 2^w - 1 > 0 &\iff \\
     (2^\lambda-1)(2^w+1) > 0
    \end{split}
  \end{equation*} 
  which is obvious. However, for proving $ \geq 2^{w-1}$, 
  \begin{equation*}
    \begin{split}
      \frac{(2^\lambda+1)(2^{w}-1)}{2^{\lambda+1}} \geq 2^{w-1} &\iff \\
     (2^\lambda+1)(2^{w}-1) \geq 2^{w+\lambda} &\iff \\
     2^{w} - 2^{\lambda} \geq 1 
    \end{split}
  \end{equation*} 
  This doesn't always hold. If the above relation doesn't hold, we can come back for $\theta = \lambda$, in this case, the relation for $\geq 2^{w-1}$ is easy to obtain. For proving $< 2^w$, we can obtain a relation $2^w - 2^\lambda < 1$ which is the contradictory side of the above relation. Since the two relation cannot hold at the same time, we can conclude that $\theta \in \{\lambda, \lambda+1\}$. For any $\theta \notin \{\lambda, \lambda +1\}$, the situation is similar to lemma \ref{lem:our_mul}. We omit the details here. Thus, lemma \ref{lem:our_add} holds as well.
\end{proof}

\section{Definition of \zkTI protocol} \label{app:D}

\begin{definition}
  \label{def:zkTI}
  The zero-knowledge truth inference protocol should satisfy the following properties completeness, soundness and zero-knowledge:  

  \begin{itemize}[leftmargin=0.3cm, itemindent=0cm]
    \item \textbf{Completeness}. For any fixed algorithm $f$ and the truth inference process $\answer^*, \quality = f(\answer, \quality)$, the proof generated by the prover is always valid, that is: 
    \begin{equation*}
      \Pr[\mathsf{zkTI.Verify}(\{\mathsf{com}_{\answer_j}\}_{j\in[m]}, \mathsf{com}_{\answer^*}, \quality, \pi, \mathsf{pp}) = 1] = 1
    \end{equation*}
    \item \textbf{Soundness}. For any PPT adversary \adversary, the following probability is $\mathsf{negl{(\lambda)}}$: 
    \begin{align*}
      &\Pr\left[
        \begin{array}{l}
          \mathsf{pp} \leftarrow \mathsf{zkTI.Setup(1^\lambda)} \\[3pt]
          (\mathsf{com}_{\answer^{*'}}, \answer^{*'}, \answer, \quality', \pi') \leftarrow \adversary(1^\lambda, \mathsf{pp}) \\[3pt]
          \mathsf{zkTI.Verify}(\{\mathsf{com}_{\answer_j}\}_{j\in[m]}, \mathsf{com}_{\answer^{*'}}, \quality', \pi', \mathsf{pp}) = 1 \\[3pt]
          \answer^{*'}, \quality' \neq f(\answer)
        \end{array}
      \right]
    \end{align*}
    \item \textbf{Zero-knowledge}. Given $x = (\{\mathsf{com}_{\answer_j}\}_{j\in[m]}, \mathsf{com}_{\answer^{*'}}, \quality)$ as the public inputs, there exists a PPT simulator $\mathcal{S}$ that for any PPT adversary \adversary, the output of the simulator is indistinguishable from the real proof. The following relation holds:
    $$
      \mathsf{View}(\adversary(\mathsf{pp}, x)) \approx \mathcal{S}^{\adversary}(x)
    $$
  \end{itemize}
\end{definition} 

\section{Proof of Theorem \ref{thm:1}} \label{app:C}

\begin{proof}  
  We give an illustration that the circuits hold in a relative large finite field $\FF$. Suppose $\delta = 2^{-(w-1)}$, to let the relations \ref{eq:our_mul}, \ref{eq:our_div}, \ref{eq:our_add} hold, we need the computation over the field not wrap around (exceed the upper bound of the field). We take relation \ref{eq:our_add} for an example. In this relation, the biggest number we need to compute is $\delta^{-1} \cdot (s_a \cdot 2^\lambda + s_b - s_c \cdot 2^\theta)$, it is easy to obtain that the upper bound of this value is $2^{3w+1}$. Thus, under a finite field where $p > 2^{3w+1}$, relation \ref{eq:our_add} holds. Then the correctness of the circuit is straightforward. Situations for the other two relations are similar.

  Next we prove the properties for Theorem \ref{thm:1} in a given relative large finite field $\FF$. For the sake of simplicity, we omit the public inputs \quality, \prior here.
  
  \noindent\textbf{Completeness.} \aggregator first run $f(\answer)$ to get the inferred result $\answer^*$ along with the satisfied extended witness \aux. Then the completeness of the protocol follows the underlying ZKP backend and the correctness of the decimal arithmetic circuits designed.

  \noindent\textbf{Soundness.} By the extractability of the ZKP backend we use, there is a PPT extractor $\mathcal{E}$ that can extract the witness $w' = \{\answer', {\answer^*}', \aux'\}$. 
  
  According to the definition of \zkTI protocol, if
  $\mathsf{com}_{\answer} = \mathsf{Commit}(\answer)$
  ,
  $\mathsf{com}_{{\answer^*}'} = \mathsf{Commit}({\answer^*}')$
  , and the \textsf{Verify} algorithm outputs 1: \\
  $\mathsf{zkTI.Verify}(\mathsf{com}_{\answer}, \mathsf{com}_{\answer^{*'}}, \pi, \mathsf{pp}) = 1$ 
  but
  $\answer^{*} \neq f(\answer) $
  , then there are two cases:
  \begin{itemize}[nosep, leftmargin=0.3cm, itemindent=0cm]
    \item Case 1. $w'$ is a valid witness for circuit \circuit. This means \aggregator could generate another group of $\aux'$ that satisfies the circuit or break the commitment scheme. the probability that \aggregator could generate such $w'$ is zero as the algorithm $f$ is a deterministic algorithm and the decimal arithmetic circuits are of security in finite field $\FF$. And the probability that \aggregator could break the commitment scheme is negligible in $\lambda$.
    \item Case 2. $w'$ is not a valid witness, i.e., $\circuit(x, w') = 0$. Then \aggregator must break the soundness of the underlying ZKP backend, the probability of which is negligible in $\lambda$.
  \end{itemize}
  Combining these two cases, the soundness of the \zkTI protocol is
  negligible in $\lambda$.

  \noindent\textbf{Zero-knowledge.} The zero-knowledge property follows directly from hiding property of the commitment scheme and the zero-knowledge property of the ZKP backend we use.
\end{proof}